\documentclass[%
reprint,
superscriptaddress,
amsmath,amssymb,
aps,
prr,
floatfix,
]{revtex4-1}

\usepackage{graphicx}
\usepackage{dcolumn}
\usepackage{bm}
\usepackage{braket}
\usepackage[german,english]{babel}
\usepackage[colorlinks=true,linkcolor=blue,citecolor=blue,breaklinks]{hyperref}
\usepackage[pdftex]{color}
\usepackage{ulem}
\usepackage{comment}
\usepackage{natbib}
\usepackage{xcolor}
\usepackage{tikzsymbols}

\begin{document}

	\title{Higher-Order Entanglement and Many-Body Invariants \\ for Higher-Order Topological Phases}
	\author{Yizhi You}
	\affiliation{Princeton Center for Theoretical Science, Princeton University, NJ, 08544, USA}
\author{Julian Bibo}
	\affiliation{Department of Physics, Technical University of Munich, 85748 Garching, Germany}
	\affiliation{Munich Center for Quantum Science and Technology (MQCST), D-80799 Munich, Germany }
\author{Frank Pollmann}
	\affiliation{Department of Physics, Technical University of Munich, 85748 Garching, Germany}
	\affiliation{Munich Center for Quantum Science and Technology (MQCST), D-80799 Munich, Germany }

	\date{\today}
	
	\begin{abstract}
We discuss how strongly interacting higher-order symmetry protected topological (HOSPT) phases can be characterized from the entanglement perspective: 
First, we introduce a topological many-body invariant which reveals the non-commutative algebra between flux operator and $C_n$ rotations. 
We argue that this invariant denotes the angular momentum carried by the instanton which is closely related to the discrete Wen-Zee response and the fractional corner charge. 
Second, we define a new entanglement property, dubbed `higher-order entanglement', to scrutinize and differentiate various higher-order topological phases from a hierarchical sequence of the entanglement structure.
We support our claims by numerically studying a super-lattice Bose-Hubbard model that exhibits different HOSPT phases. 

\end{abstract}

\maketitle

\section{Introduction}
A decade of intense effort has resulted in a thorough classification and characterization of symmetry protected topological phases of fermionic and bosonic systems~\cite{Pollmann2010,Schuch2011-jx,chen2011two,Senthil2015-tp}.
In a recent step forward, the concept of symmetry protection has been extended to include spatial symmetries~\cite{yao2010fragile,fu2011topological,hughes2011,hsieh2012topological,cheng2016translational,ando2015topological,slager2013space,hong2017topological,qi2015anomalous,huang2017building,teo2013existence,song2017topological,watanabe2017structure,po2017symmetry,isobe2015theory}.
In addition to protected gapless boundary modes, some topological crystalline phases admit gapped edges or surfaces separated by gapless corners or hinges, exemplifying a much richer bulk-boundary correspondence.
Insulators of this type are now termed higher-order topological 
insulators~(HOTIs)~\cite{benalcazar2017quantized,benalcazar2017electric,schindler2017higher,langbehn2017reflection,song2017d,song2017d,else2019fragile}.

Despite the rapid progress in the theoretical understanding of HOSPT phases (or topological crystalline phases, broadly defined)~\cite{isobe2015theory,huang2017building,song2017interaction,song2017topological,you2018higher,rasmussen2018intrinsically,rasmussen2018classification,thorngren2018gauging,benalcazar2018quantization,zhang2019construction,tiwari2019unhinging,you2018higher,jiang2019generalized}, experimentally accessible signatures or numerical fingerprints for recognizing HOSPT phases still remains challenging. 
In particular, the observation of gapless modes at the corners or hinges does not fully guarantee the bulk being HOSPT~\cite{you2019multipolar,tiwari2019unhinging}. 
Alternatively, some spatially protected SPT phases can exhibit fully gappable boundaries (including corners and hinges)~\cite{pollmann2012symmetry,liu2019shift}, while the bulk still displays a non-trivial entanglement structure which distinguishes itself from a direct product state.
The wide variety of proposals for strongly interacting HOSPT phases calls for a many-body invariant that captures their key characteristic physical phenomena~\cite{you2019multipolar}. 
A variety of topological invariants have been proposed~\cite{araki2019mathbb,zhu2019identifying} utilizing Berry phases and entanglement spectra. 
In contrast to these approaches, which are motivated by the non-interacting limit, we introduce several universal many-body invariants specifically for strongly interacting HOSPT phases. 
In particular, we propose two complementary approaches to characterize different HOSPT phases:

First, we introduce a many-body invariant that differentiates non-trivial HOSPT phases from trivial ones based on the fact that its U(1) instantons carry angular momentum, which implies that the U(1) flux insertion operator does not commute with the $C_n$ rotation symmetry. 
This non-commutative algebra uniquely characterizes HOSPT phases with fractional charges at the corners. 
One can further relate this many-body invariant to
the discrete Wen-Zee~\cite{wen1990topological,wen1998topological,liu2019shift} response, which intertwines the U(1) gauge field and spin connection.
The Wen-Zee response can be probed either by tracking the angular momentum shift under a $2\pi$ gauge flux insertion or via measuring the charge density distribution in the presence of disclinations~\cite{liu2019shift,you2018highertitus,han2019generalized,else2019fragile,else2019topological}.
Remarkably, such topological response could potentially be probed and simulated in ultracold atom systems with synthetic gauge fields created by laser-assisted tunneling or rotating traps.

Second, we propose a general recipe to detect HOSPT phases from a new `higher-order entanglement' perspective.
Different from conventional 1D symmetry protected topological phases, where the entanglement spectrum~\cite{Li-2008} displays gapless (or degenerate) modes akin to the edge spectrum~\cite{Levin-2006,KitaevPreskill,Li-2008,Pollmann2010, pollmann2012symmetry}, some HOSPT phases might exhibit a gappable (non-degenerate) and featureless entanglement spectra under any arbitrary, symmetry allowed spatial cut.
More precisely, if we merely cut out a $C_n$ wedge or perform a $C_n$ symmetric bipartition, the entanglement spectrum could display a unique ground state even if the state is in a non-trivial HOSPT phase. 
This implies that the conventional diagnosis of entanglement spectra fail to detect many HOSPT phases.
Thus the question arises whether one can still reveal fingerprints of HOSPT phases using entanglement spectroscopy. 
We introduce a new type of entanglement property, dubbed `higher-order entanglement' as a fingerprint to differentiate topological distinct HOSPT phases. 
The entanglement branching structure refers to a hierarchical sequence of entanglement spectra instead of a single spectrum. 
By symmetrically bipartitioning a $C_n$ symmetric wave function, we initially obtain the first-order entanglement spectrum which might contain non-degenerate eigenstates.
Each non-degenerate eigenstate, upon further bipartitions, should then eventually at some order exhibit a fully degenerate spectrum with respect to each $C_n$ wedge. 
Consequently, the entanglement of HOSPT phases manifests a branching structure, where any non-degenerate eigenvector of the initial entanglement spectrum contains a degenerate entanglement spectrum upon further cuts.

\section{HOSPT's in plateaus of the super-lattice Bose-Hubbard model}\label{sec-model}
\begin{figure}[t!]
\includegraphics[width=\columnwidth]{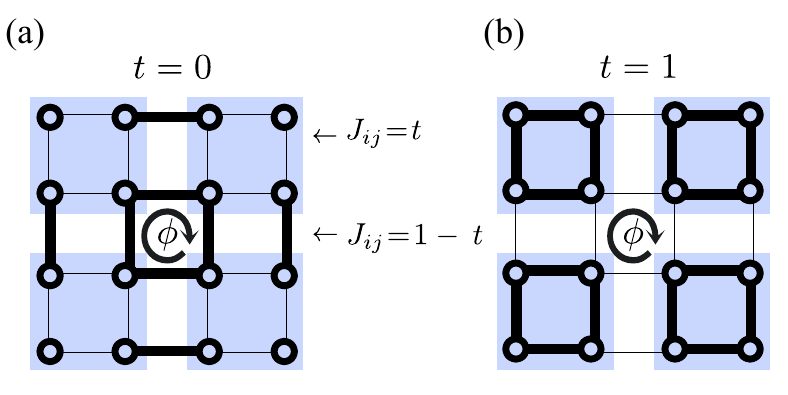}
\caption{Super-lattice hardcore boson model on a square lattice with a $2\times2$ unit cell with couplings $t$ within and $1-t$ between unit cells, respecively. Gapped HOSPT phases occurring at commensurate bulk fillings factors $n_0=1/4,1/2,3/4$ exhibiting different HOSPT orders (a) $t=0$ and (b) $t=1$. A flux insertion of $\phi$ at the central plaquette is used to define a many-body invariant (see text for details).} 
\label{fig:lattice}
\end{figure}
For concreteness, we consider a model of hardcore bosons on a $L\times L$ square lattice with a $2\times2$ unit cell. 
The Hamiltonian reads 
\begin{eqnarray}
 H &=& -\sum_{\langle i,j\rangle} \frac{J_{ij}}2(b_i^{\dag}b_j+\mathrm{h.c}) - \mu \sum_i \hat{n}_i,\label{eq:ham_b}
\end{eqnarray}
where $b_i^{\dag} (b_i)$ create (annihilate) a hardcore boson on site $i$. 
The couplings are either $J_{ij}=t$ or $J_{ij}=1-t$ with $t\ge0$ as illustrated in Fig.~\ref{fig:lattice}.
The Hamiltonian is equivalent to a super-lattice spin $S=1/2$ XY model in the presence of a magnetic field
\begin{eqnarray}
H=-\sum_{\langle i,j\rangle} \frac{J_{ij}}2(S^+_iS^-_j+\mathrm{h.c}) - \mu \sum_i S_i^z\label{eq:ham_h}, \nonumber\\
\end{eqnarray}
where $S_i^{\sigma}$, $\sigma\in\{x,y,z\}$, are the spin operators.
The Hamiltonians are $C_{4}$ symmetric with respect to the center of the lattice and preserve the total particle number (magnetization) $ N{\,}(M)=\sum_{i} n_i{\,}(S^{z}_i)$.
For $\mu =0$, the system is particle-hole (time-reversal) symmetric.
\begin{figure}
\includegraphics[width=\columnwidth]{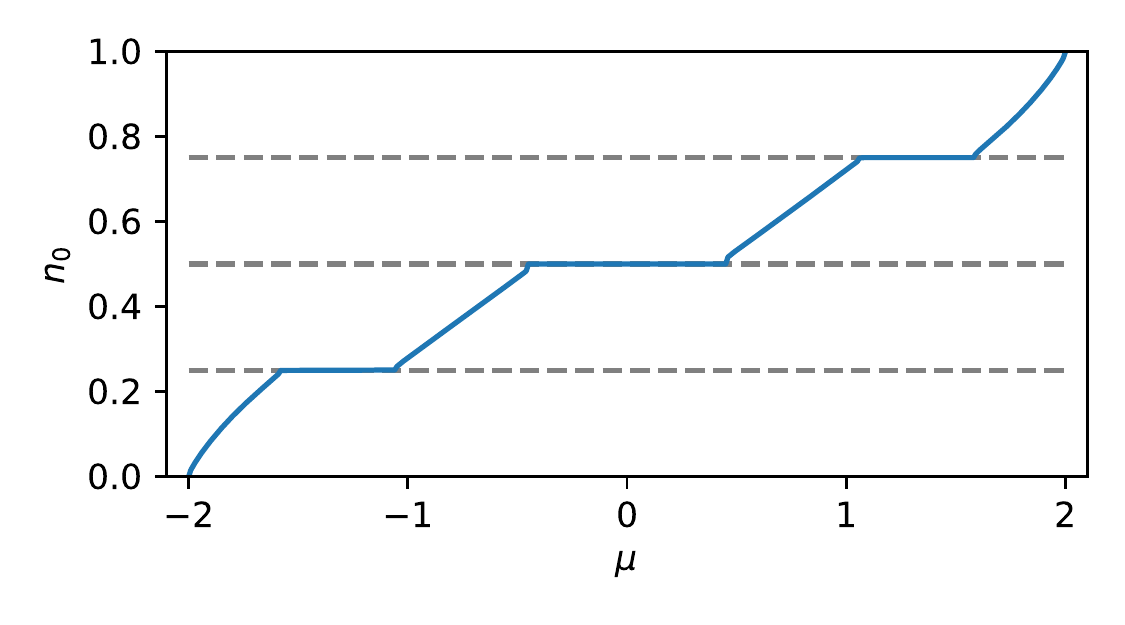}
\caption{Particle density $n_0$ as function of the chemical potential $\mu$ for the super-lattice Bose-Hubbard model obtained from DMRG simulations on an infinite cylinder with circumference $L_y = 6$ and $t=0.1$. Extended plateaux occur at commensurate bulk particle densities $n_0= 1/4,1/2,3/4$.} 
\label{fig:pd}
\end{figure}

The phase diagram of Hamiltonian~(\ref{eq:ham_b}) as function of the chemical potential $\mu$, obtained from density matrix renormalization group (DMRG)~\cite{White:1992,Hauschild2018} simulations, is shown in Fig.~\ref{fig:pd}.
The particle density shows extended, gapped plateaux separated by gapless superfluid regions.
While we show here the case $t=0.1$, an identical structure would show up for $t=0.9$ (more generally, exchanging $t\leftrightarrow 1-t$ leaves the spectral bulk properties unchanged).
The ground states in the plateaux are adiabatically connected to the zero-correlation length limits at $t=0$ and $t=1$, respectively.
In these limiting cases, either ground states can be represented as a plaquette product state $|GS\rangle=\prod_{\{\square\}}|\psi_{n_0}^{\square}\rangle$, where the product is over all plaquettes with strong bonds and $n_0$ is the average bulk filling, 
\begin{align}
&|\psi_{1/4}^{\square}\rangle= \frac12(|1000\rangle+|0100\rangle+|0010\rangle+|0001\rangle)\label{eq:pl14}\\
&|\psi_{1/2}^{\square}\rangle= \frac{1}{2\sqrt{2}}(|1100\rangle+|0110\rangle+|0011\rangle+|1001\rangle)\nonumber\\
& \ \ \ \ \ \ \ \ \ \ \ \ \ \ \ +\frac{1}{2}(|1010\rangle+|0101\rangle) \label{eq:pl12}\\
&|\psi_{3/4}^{\square}\rangle= \frac12(|0111\rangle+|1011\rangle+|1101\rangle+|1110\rangle)\label{eq:pl34}.
\end{align}

The difference between the cases $t=0$ and $t=1$ becomes clear when considering clusters with open boundary conditions. 
In the \textit{trivial phase}, $t=1$, the ground states are unique for all plateaux (see~Fig.~\ref{fig:lattice}b). 
In the \textit{topological phase}, $t=0$, the four corners are decoupled and each of them can be either filled or empty (see~Fig.~\ref{fig:lattice}a).
Let us consider the case $n_0=1/2$ in detail:
At fine tuned $\mu=0$, we find $2^4=16$ degenerate states depending on the occupancies of the four corners -- zero modes are formed at the corners.
If $\mu$ is detuned from zero, we obtain a unique $C_4$ symmetric ground state for which the total particle number deviates from exact half-filling by a filling anomaly of $\pm 2$.
Note that the degeneracy at the corner is no longer protected when particle-hole (time-reversal) symmetry is broken. 
The filling anomaly gives rise to quantized fractional corner charges which are protected by the $C_4 \times U(1)$, as it was already discussed in the context of HOTI's~\citep{benalcazar2017quantized}. 
Comparing the charge distributions, we find a fractional charge $Q_{\rm corner}=1/2$ localized around the corners in the topological phase, measured with respect to the average bulk filling~\cite{bibo2019fractional}.
Analogously, we obtain a similar picture for filling factors $n_0=1/4 $ and $n_0=3/4$ having fractional corner charges $Q_{\rm corner}=1/4$ and $Q_{\rm corner}=3/4$ for open boundaries, respectively.
As in the half-filled case, the fractional corner charges are a consequence of the filling anomalies $\pm 1$ and $\pm 3$, respectively. 
%
Note that, unlike to the half-filled case, we have to add additional, symmetry preserving terms to the Hamiltonian to guarantee gapped edges with appropriate filling factors.
Despite the existence of fractional corner charges, the gapless modes at the corners can be symmetrically gapped out by turning on the Zeeman field at each corner to pin the particle configurations. 
Subsequently, the degeneracy at the corner is merely a consequence of the filling anomaly and is not protected by symmetry. 

In the following, we will derive bulk invariants that can be used to characterize the different gapped HOSPT phases. 

\section{Many-body invariant for HOSPT phases}
Despite the fact that the mathematical structure and classification of interacting HOSPT phases (or topological crystalline phases, broadly defined) is now well understood \cite{isobe2015theory,song2017interaction,song2017topological,rasmussen2018classification,you2018higher,rasmussen2018intrinsically,rasmussen2018classification,thorngren2018gauging,tiwari2019unhinging,you2018higher}, experimentally accessible bulk many-body invariants \cite{kang2018many,wheeler2018many,araki2019mathbb} for the characterization of such phases are still lacking. 
The obstacle lies in the fact that the associated topological or entanglement structure cannot be measured by any local operator. In this section, we introduce a many-body invariant for HOSPT phases, which is closely related to the Wen-Zee response~\cite{wen2003quantum,witten1991quantization}, where a flux insertion changes the angular momentum in specific ways, and leverage its relationship with the fractional corner charge. 
For 2D SPT phases, that are protected by internal symmetries, their topological response can be categorized as the Chern-Simons type, where a flux excitation either contains a projective symmetry or carries a charge~\cite{wen1990topological}. 
Hence, after a symmetry gauging procedure, the resulting gauge flux carries fractional statistics or projective zero modes. 
When it comes to a spatial $C_n$ rotation symmetry, under coarse-graining or renormalization toward the long wave-length limit, the $C_n$ symmetry on the lattice can be treated as an internal $Z_n$ symmetry at IR and the rotation of the lattice is dual to the internal permutation of the $Z_n$ boson living inside the enlarged unit cell. 
Such IR mapping between $C_n$ rotation and internal symmetry is well-established for the classification and characterization of topological crystalline phases.
To be more specific, any HOSPT phase protected by $C_n$ rotation symmetry can be traced back to an SPT with $Z_n$ symmetry~\cite{you2018highertitus,thorngren2016gauging,thorngren2018gauging}.

\subsection{Flux insertion and Wen-Zee response}
Let us now return to the model introduced in Sec.~\ref{sec-model} with $C_4\times U(1)$ symmetry.
Based on our previous argument, a HOSPT phase protected by $C_4\times U(1)$ symmetry can be traced back to the $Z_4\times U(1)$ SPT state characterized by the mutual Chern-Simons response. 
In particular, a $U(1)$ flux insertion could trap a discrete $Z_4$ charge, hence the U(1) flux operator does not commute with $Z_4$ symmetry. 
By replacing the $Z_4$ symmetry with the spatial $C_4$ symmetry, the $Z_4$ quantum number becomes the angular momentum modulo four and an instanton event (i.e., the U(1) flux insertion) changes the spatial $C_4$ eigenvalue.


To be more explicit, we introduce the instanton event, which can be entitled as a flux insertion operator,
\begin{align}
U_{2\pi}=e^{i 2\pi \theta \hat{n}(r,\theta)}
\end{align}
with $\theta$ being the polar angle with respect to polar coordinates and $\hat{n}$ being the density operator.
The operator $U_{2\pi}$ introduces a $2\pi$ flux at the central plaquette as shown in Fig.~\ref{fig:lattice}. As a consequence, the boson hopping around the central plaquette gets a phase modulation. 
Before flux insertion, the ground state carries zero angular momentum (modulo 4). 
After flux insertion, the plaquette entangled wave function changes its sign structure and subsequently the total angular momentum is shifted by $l=1,2,3$ depending on the magnetization. 
To compare the angular momentum shift after flux insertion, we calculate the commutation relation between the flux operator and the $C_4$ symmetry,
\begin{align}
U_{2\pi}C_4= e^{i \frac{\pi}{2} (\sum_i \hat{n}_i) }C_4 U_{2\pi}=e^{i \frac{\pi}{2} N }C_4 U_{2\pi}
\label{xiechengoddess}
\end{align}
The commutation relation depends on the total charge modulo 4 which is equivalent to the magnetization or fractional charge density at each corner.
When the total boson number is $4k+l$ with $l/4$ charge per quadrant \footnote{We can lift the corner degeneracy by adding a magnetic field at the corners resulting in fully gapped edges and corners, respectively. Subsequently, due to the plaquette entangled state at the center, each quadrant contains a fractional charge $l/4$ .}, the flux insertion operator does not commute with $C_4$ rotation symmetry, hence the
the angular momentum is shifted by $l$ after flux insertion~\cite{liu2019shift}. 
This angular momentum shift with respect to the U(1) flux is universal for any HOSPT system regardless of the microscopic form of the Hamiltonian and is merely determined by the fractional corner charge at each quadrant. 

In general, the angular momentum response with respect to flux insertion can be described by the Wen-Zee response~\cite{wen1998topological,liu2019shift},
\begin{align}
\frac{l}{2\pi}\omega \wedge dA
\label{wz}
\end{align}where $\omega$ is the discrete version of the spin connection. 
The curl of the spin connection $(\partial_x \omega_y-\partial_y \omega_x)$ gives the disclination flux, which is exactly the symmetry flux of the $C_4$ rotation symmetry. 
The mutual coupling between spin connection and U(1) gauge field implies that the $2\pi$ instanton carries a $C_4$ rotation charge $l{\,}\rm{mod}{\,}4$. 
Alternatively, if we gauge the $C_4$ symmetry by inserting a $\pi/2$ disclination flux, the resultant charge density trapped inside the disclination core is $\rho=l/4$, which exactly matches the corner charge~\cite{you2018highertitus}. Physically, the disclination is generated by removing a quadrant and reconnecting the boundary, such that each disclination center contains a fractional charge density equal to the corner charge.
The angular momentum shift with respect to the flux insertion can only take discrete values~\cite{liu2019shift}, resuling in a level quantization of the Wen-Zee term, which is expected for gapped, short-ranged entangled systems. Remarkably, we expect that the invariant can be probed in cold atom or ion trap experiments by introducing an artificial U(1) gauge flux created by rotating traps or coherent light–matter interaction and measuring the angular momentum shift implemented by local random unitaries~\cite{elben2019many,Enk2012,Elben2018}.

\subsection{Numerical measurement of Wen-Zee response}
\begin{figure}[t!]
\includegraphics[width=\columnwidth]{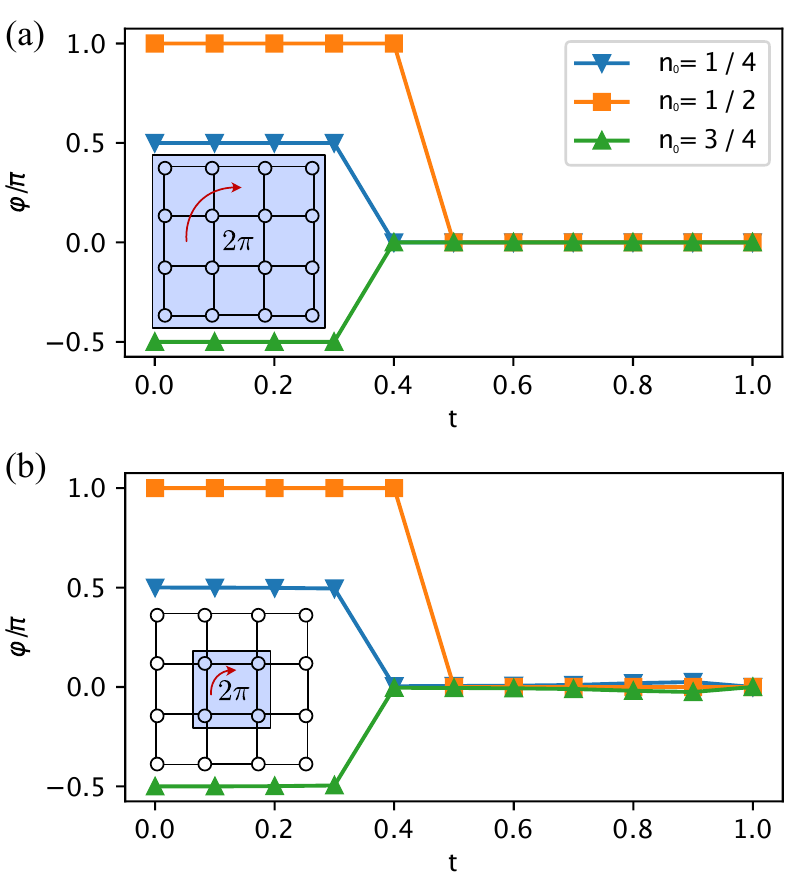}
\caption{Angular momentum shift of Hamiltonian (\ref{eq:ham_b}) after inserting a $2\pi$ flux through the central plaquette. The numerical data is shown for $4\times 4$ clusters at different commensurate filling factors $n_0=1/4,1/2,3/4$ with gapped corners. Panel (a) shows the response for a $\pi/2$ rotation of the full lattice and (b) for a partial rotation that only involves the center plaquette.} 
\label{fig:flux}
\end{figure}

We now numerically compute the previously introduced many-body invariant using exact diagonalization of small clusters (see Fig.~\ref{fig:flux}).
For this we compare the angular momentum quantum number $e^{i\varphi} = \langle\psi(\phi)|C_4|\psi(\phi)\rangle$ with $\varphi = \frac{l\pi}{2}{\,}{\rm mod}{\,}2\pi$ and $l=1,2,3$ of the ground state with flux $\phi=0$ and $\phi=2\pi$ for clusters with gapped corners.
While we always find $\varphi=0$ for $|\psi(\phi=0)\rangle$, the angular momentum of $|\psi(\phi=2\pi)\rangle$ differs between different HOSPT phases. 
To illustrate the results, let us consider the case $t=0$ at filling $n_0=1/4$ without flux for which the center plaquette has the simple form
\begin{align}
&|\psi^{\square}_{1/4}(\phi=0)\rangle= \frac12(|1000\rangle+|0100\rangle+|0010\rangle+|0001\rangle).\nonumber
\end{align}
We find $\langle\psi_{1/4}(\phi=0)|C_4|\psi_{1/4}(\phi=0)\rangle=1$ and thus this state has zero angular momentum. 
After inserting a flux of $2\pi$, the ground state is 
\begin{align}
&|\psi^{\square}_{1/4}(\phi=2\pi)\rangle= \frac12(|1000\rangle+i|0100\rangle-|0010\rangle-i|0001\rangle),\nonumber
\end{align}
with $\langle\psi_{1/4}(\phi=0)|C_4|\psi_{1/4}(\phi=0)\rangle=i$ and the state has an angular momentum shift of $l=1$. 
Analogously, we can understand the response for filling factors $n_0=1/2$ and $n_0=3/4$. 
As the quantized Wen-Zee response in Eq.~(\ref{wz}) is local, a $2\pi$ flux insertion changes the wave function configuration around the center only within an area spanned by the correlation length.
Next we consider the response due to a partial rotation of a symmetric block of sites around the center (e.g., the central plaquette $\tilde{C}_4$).
Since $\tilde{C}_4$ does not commute with the Hamiltonian (except in the limiting case $t=0$ and $t=1$), we find $\langle\psi(\phi)|\tilde{C}_4|\psi(\phi)\rangle \sim e^{-\gamma \tilde{\ell}_{B}}e^{i\varphi}$, where $\tilde{\ell}_{B}$ is the linear size of the rotated block and $\gamma>0$ some constant.
The angular momentum shift extracted from this quantity can still be used to characterize the phases as shown in Fig.~\ref{fig:flux}b.
Since the partial rotation can be obtained using randomized measurements, implemented with local random unitaries~\cite{elben2019many,Enk2012,Elben2018}, it is expected to be accessible in any spin system with single-site addressability and readout.

\section{Entanglement diagnosis for HOSPT phases}

The study of many-body entanglement, as obtained from the reduced density matrix $\rho_A$ for a bipartition of the system into two disjoint parts $A$ and B, has been shown to be a very useful tool for the characterization of quantum phases of matter~\cite{RevModPhys.80.517}.
Moreover, the relation between the topological structure and the entanglement spectrum, i.e., the spectrum of the reduced density matrix $\rho_A$, has been widely explored \cite{Li-2008,Peschel_2009,pollmann2012symmetry,PhysRevB.84.195103,PhysRevLett.105.115501,PhysRevLett.104.130502,chandran2014universal,PhysRevLett.110.236801}.
Remarkably, most salient topological properties including quasiparticle statistics, edge excitations, central charge and topological Berry phase can be readily reached by scrutinizing the entanglement spectrum.
In Ref.~\cite{zhu2019identifying} it has been proposed that certain HOSPT phases can be characterized by the entanglement spectrum, more precisely, it was suggested that the low-lying eigenvalues $e_{\alpha}$ of entanglement Hamiltonian $\mathcal{H}$ (i.e., the logarithm of the reduced density matrix) reflects the energy spectrum of the in-gap states, and hence can be treated as a fingerprint of the topological phases.

However, such straightforward correspondence between bulk topology and entanglement spectrum might not apply to strongly interacting HOSPT states. First and foremost, some interacting HOSPT states contain a featureless gapped entanglement spectrum, equivalently to their trivial phase counterparts.
In addition, the correspondence between the low-lying part of the entanglement spectrum and the bulk topology
cannot be taken too literally~\cite{chandran2014universal}. 
Since the reduced density matrix is the partition function of the entanglement Hamiltonian (EH) at finite temperatures, the high energy modes in the entanglement spectrum (ES) also contribute to the intertwined features of the ground state. In particular, the low-lying states of the ES may undergo a phase transition while the bulk phase remains unchanged~\cite{chandran2014universal}.

\begin{figure}[ht!]
\includegraphics[width=0.9\columnwidth]{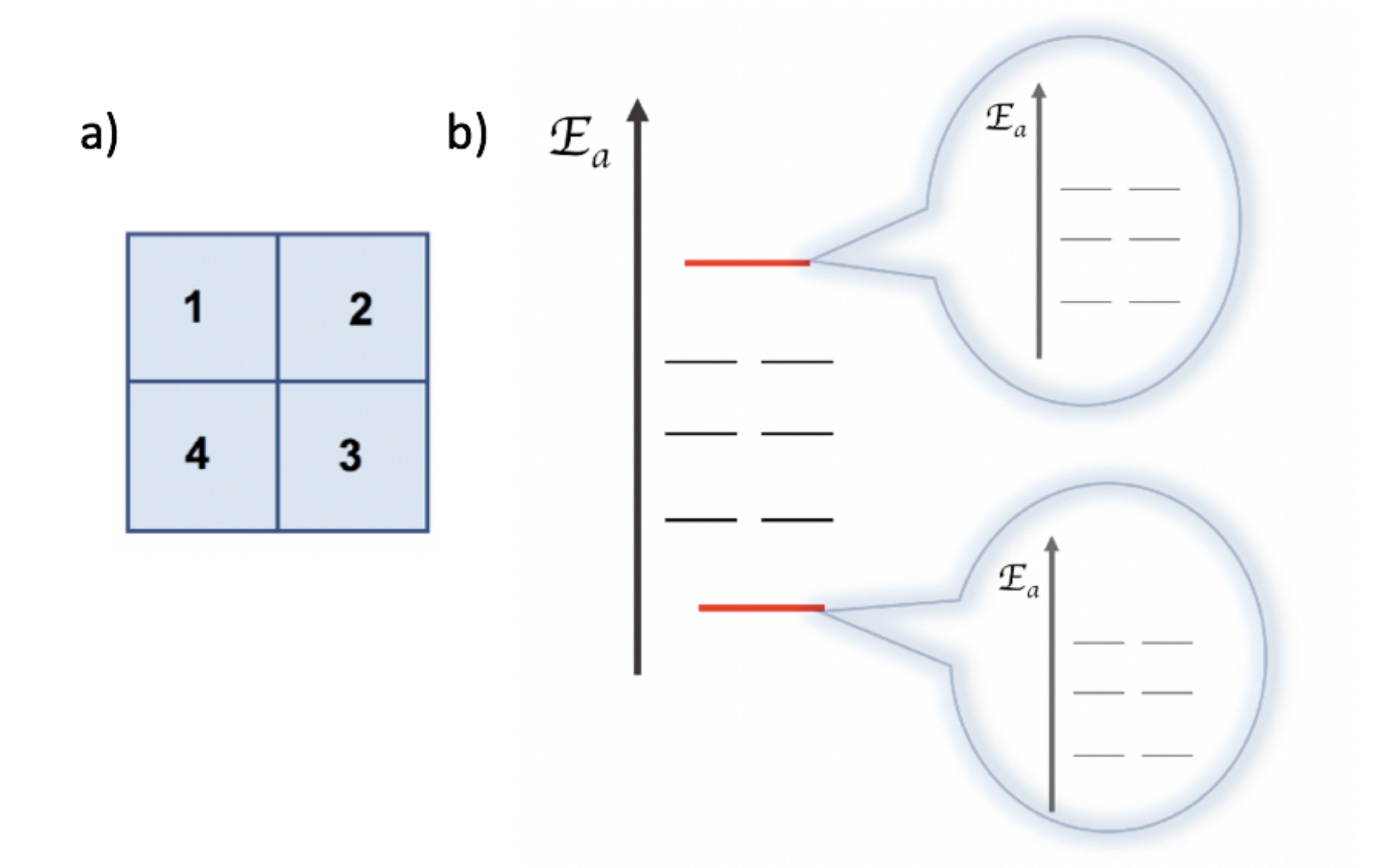}
\caption{Higher-order entanglement branch structure. In panel (a) the ground state is divided into four symmetric regions. (b) The hierarchical sequencing of the entanglement spectrum is shown. First, we trace out region ($1$-$3$) to obtain the reduced density matrix for region ($2$-$4$). Second, for any single-valued eigenvector of the entanglement Hamiltonian the entanglement spectrum between regions $2$ and $4$ is calculated showing a two-fold degeneracy for the full spectrum.} 
\label{fig:branch}
\end{figure}

We analyze the universal features of the many-body EH in various interacting HOSPT phases. It is worthy to emphasize that both the low-lying states and the highly excited part of the ES are responsible for the ground state pattern of HOSPT phases, so there is no reason to overlook the excited states in the EH. 
To set the stage, we will first establish a Kramers theorem for the EH in HOSPT phases: \textit{If the symmetry operator acting on each $C_n$ corner is projective, then the reduced density matrix with respect to each $C_n$ corner cut exhibits level degeneracies for the whole entanglement spectrum.}
However, for generic $C_n\times U(1)$ symmetric HOSPT phases, which do not render a projective symmetry at the corner, the ES upon spatial bipartition might be non-degenerate and hence cannot be treated as a fingerprint for HOSPT states. 
To conquer this obstacle, we introduce a new entanglement property -- `higher-order entanglement' as illustrated in Fig.~\ref{fig:branch}b. 
The basic idea is that we implement further bipartitions for the non-degenerate part of the spectrum, which in turn shows degeneracies. 
This higher order entanglement branching phenomena is a unique feature of HOSPT phases and is closely connected to the fractional corner charge and Wen-Zee response. 
In particular, the higher-order entanglement indicates that the traditional ES is not adequate for characterizing the topological feature of the ground state.
A complete viewpoint of the ground state structure requires a hierarchical sequence of the entanglement branch. 

In addition to the exploration of entanglement features for the ground state, we also demonstrate that the spectrum of the EH combined with the actions of the symmetries on its eigenstates is sufficient to predict the response of the phase to flux insertion, linking the entanglement spectrum to a well-known class of quantized response functions such as Wen-Zee response or rotation symmetry gauging. 
While these responses have previously been discussed elsewhere \cite{else2019topological,liu2019shift,you2018higher}, we find it useful to discuss them in the common language of entanglement, both in order to better understand the universality of the EH and for practical purposes because these responses can then be measured from entanglement information readily available using DMRG.

\subsection{Entanglement spectrum for HOSPT with projective symmetry at the corner}

\begin{figure}[ht!]
\includegraphics[width=\columnwidth]{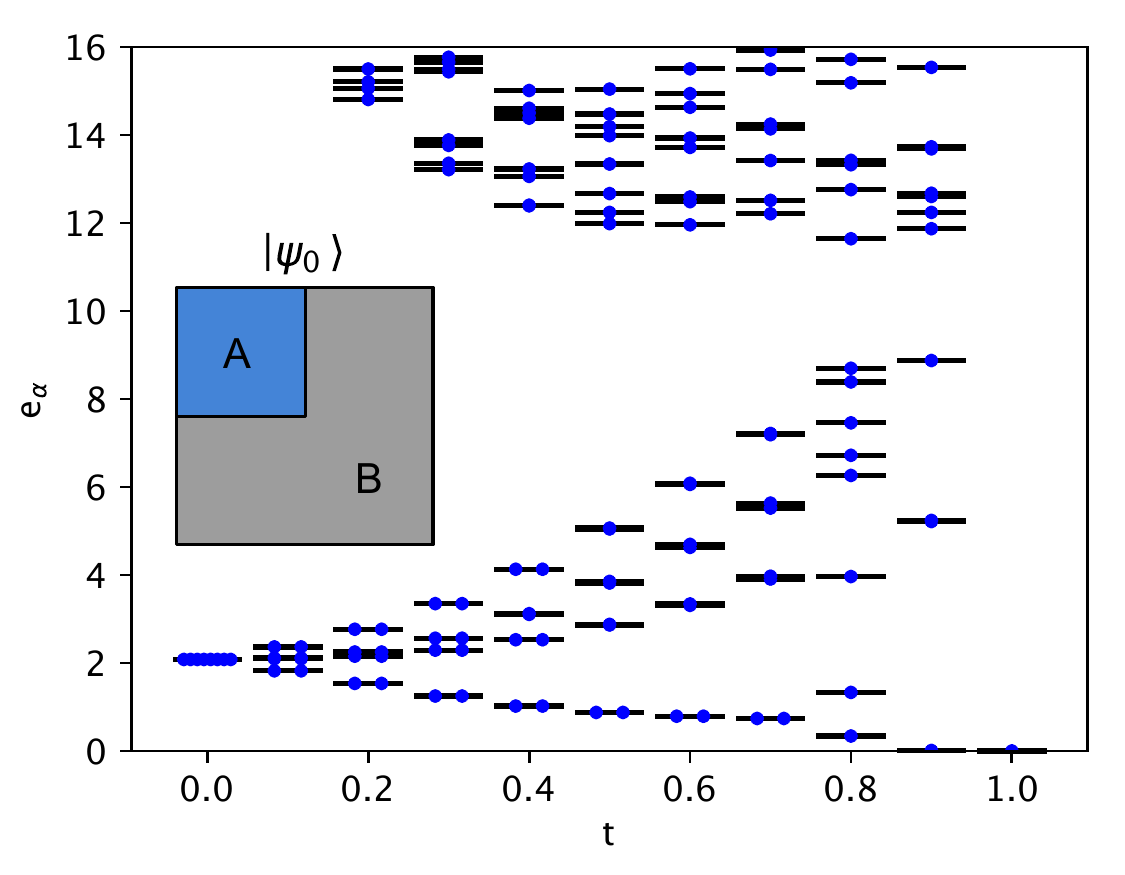}
\caption{Corner entanglement spectrum for the particle-hole symmetric case ($\mu=0$) with even degeneracies of the low-energy part in the HOSPT phase with half-charge corner states. The weak symmetry breaking potential at the corners, which we use to obtain a unique ground state, leads to a breaking of degeneracies at higher entanglement energies. In the trivial phase, we obtain a gapped featureless entanglement spectrum.} 
\label{fig:CES}
\end{figure}
In this section we explore the ES of HOSPT phases protected by $C_n \times G$ symmetry, in which $G$ renders a projective representation at the $C_n$ corner.
For example, the HOSPT phase in the super-lattice XY model with $\mu=0$ has $S=1/2$ spins localized at the corners, which are protected by time-reversal $\mathcal{T}^2 = -1$  and $C_4$ symmetry, respectively.
We denote this as the generalized `Kramers theorem' of the entanglement Hamiltonian.

Let us consider a Schmidt decomposition that cuts one quadrant out of the ground state $|\mathrm{GS}\rangle$,
\begin{align}
|\mathrm{GS}\rangle = \sum_{\alpha}\Lambda_{\alpha} |A_{\alpha}\rangle_{1}|B_{\alpha}\rangle_{2,3,4},
\end{align}
where the quadrants are defined as in Fig.~\ref{fig:branch}a and $\Lambda_{\alpha}$ are the Schmidt values. 
The Schmidt states $|A_{\alpha}\rangle_{1}$ and $|B_{\alpha}\rangle_{2,3,4}$ form an orthogonal basis of the two parts, respectively.
The reduced density matrix is diagonal in the Schmidt basis $\rho_1 = \sum_{\alpha}\Lambda_{\alpha}^2 |A_{\alpha}\rangle_{1}\langle A_{\alpha}|_{1}$ and the entanglement spectrum is given by $e_{\alpha}=-2\log\Lambda_{\alpha}$.
We first demonstrate that any Schmidt decomposition should be block diagonal in the $G$ basis, provided $G$ is a symmetry of the ground state. 
Let us act with the symmetry operation $G$ on the ground state such that
\begin{align}
G|\mathrm{GS}\rangle = \sum_{\alpha}\Lambda_{\alpha} G|A_{\alpha}\rangle_{1}G|B_{\alpha}\rangle_{2,3,4},\end{align}
As $G$ is an internal symmetry, it acts on the two regions independently. 
We then set $G|A\rangle_{1}$ as the new basis for the Schmidt decomposition and by doing so the reduced density matrix for region $1$ is,
\begin{align}
&\rho_1=
\sum_{\alpha} G \Lambda^2_{\alpha}|A_{\alpha}\rangle_1 \langle A_{\alpha}|_1 G^{-1}=G \rho_1 G^{-1}.
\end{align}
This implies that $G$ commutes with the reduced density matrix.
Now assume $G$ has a projective representation at the corners. 
We start with a simple example where $G=\mathcal{T}$ with $\mathcal{T}^2=1$ on-site. 
This is exactly the case in Eq.~(\ref{eq:ham_h}) at $\mu=0$ in the absence of a Zeeman field. 
As the reduced density matrix $\rho_1$ commutes with $\mathcal{T}$, we can regard the EH $\rho_1=e^{-H_1}$ as a many-body system with $\mathcal{T}$ symmetry. 
Since $\mathcal{T}$ is projective for each corner, we have $\mathcal{T}^2=-1$ for $\rho_1$ which indicates $\mathcal{T}$ is projective when acting on the reduced density matrix with respect to the corner region. 
Consequently, all eigenstates in the EH come in Kramer pairs.
To demonstrate this statement, we take the model in Eq.~(\ref{eq:ham_h}) at the $\mathcal{T}$ symmetric point $\mu=0$. 
In the HOSPT state ($t\rightarrow0$), the corner contains a free spin-1/2 mode with a two-fold level degeneracy resulting in a projective representation of the particle-hole symmetry at the corner.
In numerical simulations, these zero energy states at the corner would unavoidably entangle each other due to finite size effects.
To avoid such long range entanglement from the corner zero modes, we apply a weak local chemical potential at each corner to pin the corner configurations without affecting the bulk. 
Although $\mathcal{T}$ symmetry is weakly broken near the corners, the two-fold degeneracy of the low-lying states in the ES, contributed from the local entanglement near the cut-center, still persists. However, the degeneracy of the highly excited spectrum is slightly lifted as a consequence of the weak symmetry breaking near the corners. 
In Fig.~\ref{fig:CES}, we plot the ES with respect to the relative hopping amplitude $t$. 
In the HOSPT phase, the low-lying part of the ES exhibits a robust two level degeneracy. As already mentioned, the high energy part displays level splitting due to $\mathcal{T}$ symmetry breaking at the corners. 
In the trivial phase, the ES is featureless with a unique ground state.

\subsection{Higher Order Entanglement in HOSPT Phases}
\begin{figure}[ht!]
\includegraphics[width=\columnwidth]{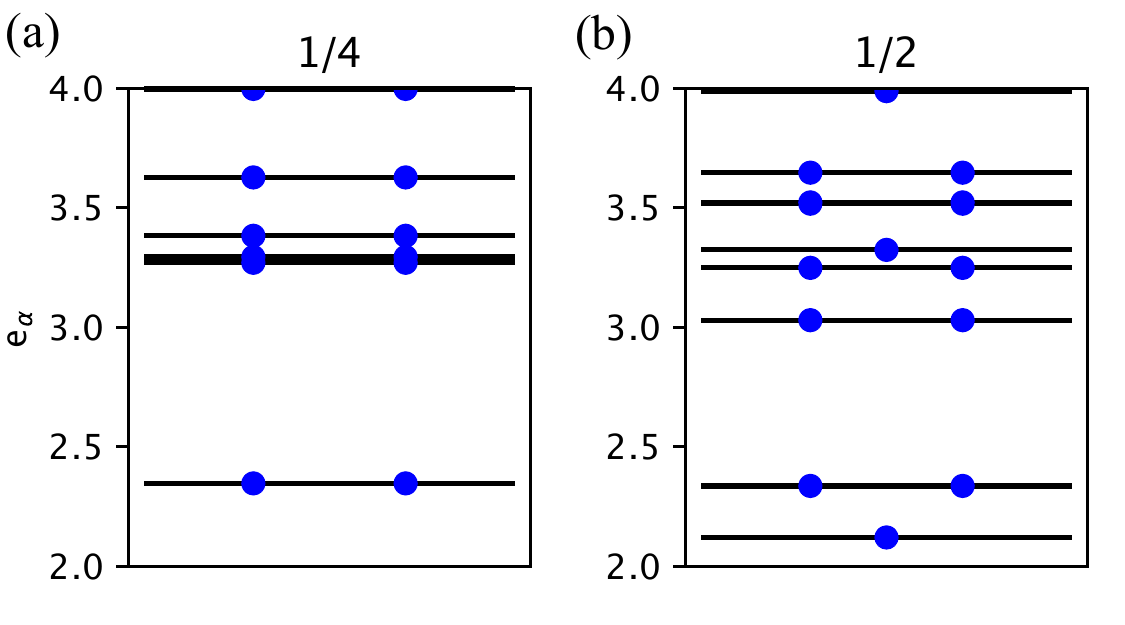}
\caption{Higher order entanglement spectrum for a $C_4$ symmetric bipartition of the ground state of Hamiltonian (\ref{eq:ham_b}) by tracing out the two diagonal corners. The simulation is made for $t=0.1$ with corner charges $1/4$ (a) and $1/2$ (b). The latter case shows a featureless entanglement spectrum with a unique ground state.}
\label{fig:es_sym}
\end{figure}
We will now explore the properties of the entanglement spectrum of HOSPT phases that do not exhibit projective representations at the corners. 
We begin by demonstrating that the HOSPT model in Eq.~(\ref{eq:ham_b}) with $C_4\times U(1)$ symmetry can host fully gappable entangled spectra for any spatial cut away from $\mu=0$ (i.e., broken particle-hole symmetry).

Let us focus on the reduced density matrix with respect to the $C_4$ symmetric quadrant cut in Fig.~\ref{fig:branch}a.
For HOSPT phases with corner charge $Q_{\rm corner}=1/4{\,}(3/4)$, away from the points with fine tuned $t$, the ES for region $1$ has a unique ground state. 
This non-degenerate spectrum is a consequence of the broken particle-hole symmetry. Thus the configurations with even or odd number of charges in each quadrant have different weights. 
Let us consider a typical fixed-point wave function of the HOSPT phases with $1/4$ charge at the corner, which can be written as a product of plaquette entangled states, analogously to Eq.~(\ref{eq:pl14}).
If we trace out a corner site from the plaquette, the reduced density matrix has unique eigenvalues $1/4$ and $3/4$, respectively.
In the meantime, if we make a $C_4$ symmetric cut by tracing out the region ($1$-$3$), we find a robust degeneracy and all eigenvalues appear in pairs as shown in Fig.~\ref{fig:es_sym}a.
If we, however, consider a HOSPT phase with corner charge $Q_{\rm corner}=1/2$, the ES for region ($2$-$4$) has unique and degenerate low-lying states (see Fig.~\ref{fig:es_sym}b).
To summarize, the traditional ES is insufficient for recognizing generic HOSPT phases since it provides only limited information about the ground state topology.
We therefore introduce a new entanglement property -- \textit{higher-order entanglement} with a hierarchical sequence of the ES to categorize distinct HOSPT phases.

\subsubsection{$C_4\times U(1)$ symmetry}
We elaborate the power of higher-order entanglement branch for HOSPT phases with $C_4 \times U(1)$ symmetry but the argument can be generalized to any $C_{2^n}$ symmetry. 
As a starting point, we bipartite the system into two $C_4$ symmetry related regions ($1$-$3$) and ($2$-$4$), respectively (see Fig.~\ref{fig:branch}a). The Schmidt decomposition for the two regions is given by
  \begin{align}
|\mathrm{GS}\rangle=\sum \Lambda_{\alpha}|A_{\alpha}\rangle_{1,3} |B_{\alpha}\rangle_{2,4}.
  \end{align}
The two halves are related by symmetry and thus the Schmidt states are transformed into each other $|A_{\alpha}\rangle_{1,3} \leftrightarrow C_4 |B_{\beta}\rangle_{2,4}$.
Since the ground state $|\mathrm{GS}\rangle$ is $C_4$ symmetric, no matrix elements connecting states with different Schmidt eigenvalues $\Lambda_{\alpha}$ can occur. 
A unique Schmidt value $\Lambda_{\alpha}$ implies that $|A_{\alpha}\rangle_{1,3} = C_4 |B_{\alpha}\rangle_{2,4}$ with two equal configurations.
When $|A_{\alpha}\rangle_{1,3} \ne C_4 |B_{\alpha}\rangle_{2,4}$, the Schmidt spectrum has to be degenerate to ensure that the state is $C_4$ symmetric (i.e., the reduced density matrix is invariant under $C_4$ symmetry).
Moreover, as the theory is $U(1)$ symmetric, each Schmidt state has a well defined charge number.

When the HOSPT phase contains fractional corner charges $Q_{\rm corner }= 1/4{\,}(3/4)$, the total charge number of the ground state is odd $4N+1{\,}(4N+3$). 
Under this circumstance, the Schmidt states $|A_{\alpha}\rangle_{1,3}$ and $|B_{\alpha}\rangle_{2,4}$ must have different $U(1)$ charges to ensure that the total charge density is odd. 
This in turn implies that $|A_{\alpha}\rangle_{1,3}\ne C_4|B_{\alpha}\rangle_{2,4}$ and thus all Schmidt values $\Lambda_{\alpha}$ must be degenerate for a $C_4$ symmetric $|\mathrm{GS}\rangle$.

When the HOSPT phase contains fractional corner charges $Q_{\rm corner }=1/2$, with the total charge being even $4N+2$, it is possible to have unique Schmidt values.
For this specific case, the ground state wave function of the $C_4 \times U(1)$ symmetry protected HOSPT phase is adiabatically connected to two AKLT chains along the diagonal and off-diagonal direction, crossing in the symmetry center. Although the corner still carries fractional charge, there is no entanglement between regions ($1$-$3$) and ($2$-$4$) resulting in a single-valued ES.

This is where the higher-order entanglement becomes crucial:
We first separate the Schmidt spectrum into a degenerate part (with $|A_{\alpha}\rangle_{1,3}\ne C_4|B_{\alpha}\rangle_{2,4}$) and a unique part (with $|A_{\alpha}\rangle_{1,3} = C_4|B_{\alpha}\rangle_{2,4}$) as shown in Fig.~\ref{fig:branch}b.
Let us further bipartite each unique Schmidt state to obtain a hierarchical sequence of the higher-order ES,
\begin{align}
&|A_{\alpha}\rangle_{1,3}=
\sum_{\gamma} \Lambda_{\gamma} |C_{\gamma}\rangle_{1} |D_{\gamma}\rangle_{3}.
\end{align}
Note that equivalently we could have considered the corresponding state $|B_{\alpha}\rangle_{2,4}$.
The two halves are related by $C_2=C_4^{2}$ symmetry and consequently the Schmidt states are transformed into each other $|C_{\gamma}\rangle_{1} \leftrightarrow C_2 |D_{\delta}\rangle_{3}$ with $\Lambda_{\gamma}=\Lambda_{\delta}$.
Moreover, all unique Schmidt states $|A_{\alpha}\rangle_{1,3}$ must have charge number $2N+1$ to guarantee the total charge of the ground state is $4N+2$. 
Thus we can use a similar argument to show that all Schmidt values must come in degenerate pairs. 
In particular, there cannot be a single state in the decomposition with $|C_{\gamma}\rangle_{1} = C_2 |D_{\gamma}\rangle_{3}$ and thus the entire spectrum must be degenerate. 
The argument flows as follow: If there is a single state with $|C_{\gamma}\rangle_{1} = C_2 |D_{\gamma}\rangle_{3}$, then $C_{\gamma}$ should carry $1/2$ charge number to guarantee the total charge being odd for $A_{\alpha}$. 
As the elementary charge is an integer, half charges only appear as the cat state $|0\rangle+|1\rangle$ breaking $U(1)$ symmetry. 
\begin{figure}[ht!]
\includegraphics[width=\columnwidth]{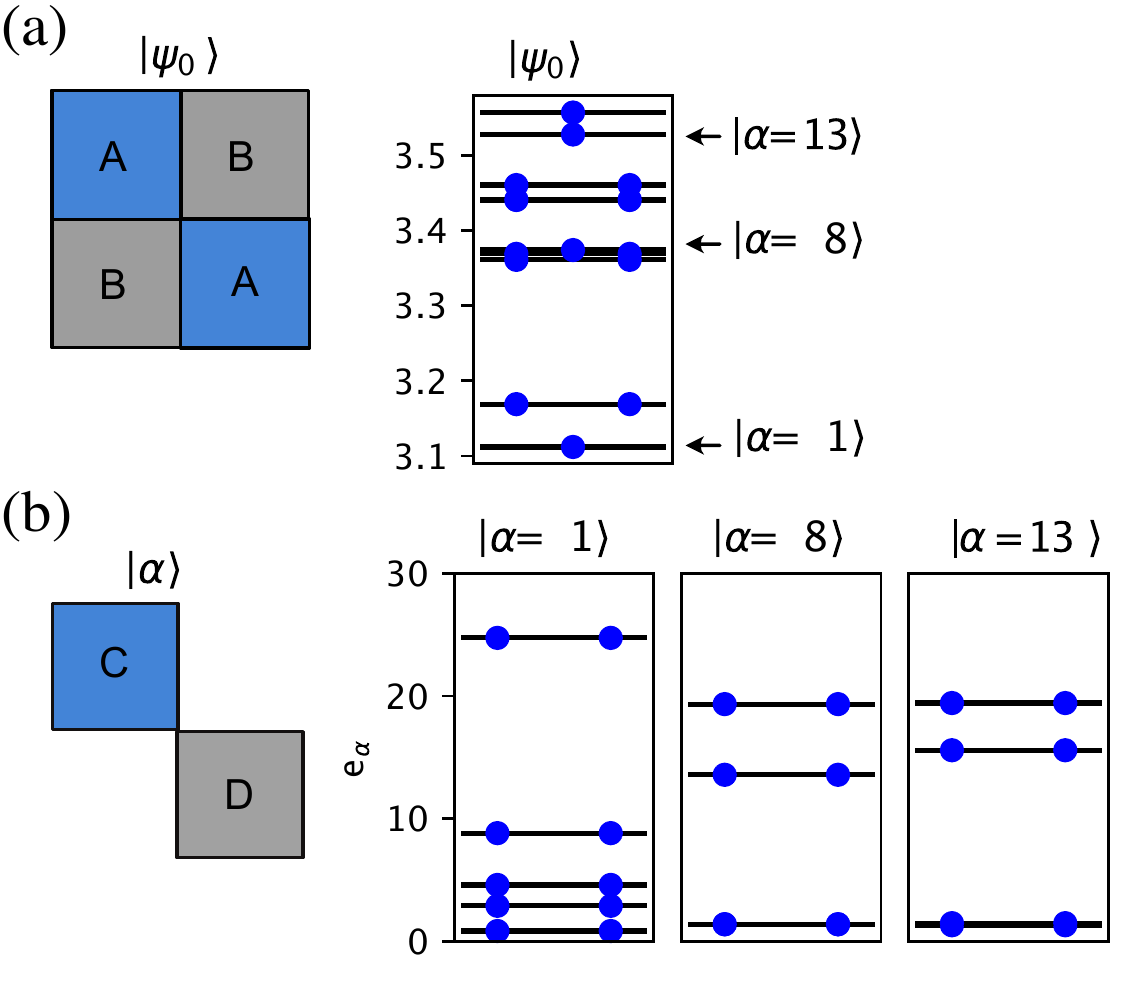}
\caption{Entanglement spectra of the HOSPT phase with half-corner charge at $t=0.1$: (a) Entanglement spectrum for a bipartition in which the two parts are related by the $C_4$ symmetry. (b) Higher-order entanglement branches of the unique Schmidt states with exact degeneracies.} 
\label{hier1}
\end{figure}

The numerical data obtained for the super-lattice Bose Hubbard model nicely shows the expected higher-order entanglement branch structure as shown in Fig.~\ref{hier1}.

To summarize, the ES of HOSPT phases contains a branching structure in which non-degenerate Schmidt states of the initial ES contain a fully degenerate higher-order ES upon further symmetric cuts.

A similar argument holds for $C_{2^n}$ symmetric HOSPT phases: 
The entanglement branching structure is similar but we just need to duplicate the bipartition step further as the initial Schmidt spectrum might contain a series of non-degenerate eigenstates. 
If we take such state out and redo the bipartition $n-1$ times, the resultant $n$-th order spectrum should always contain two-fold degeneracies. 
In particular, it is worth emphasizing that if we merely make a $C_{2^n}$ symmetric corner cut and calculate the ES of that region, the ES may not display any robust degeneracy. 
This is obvious for the plaquette entangled state $|\psi_{3/4}^{\square}\rangle$ defined in Eq.~\eqref{eq:pl34} with ES eigenvalues $1/4$ and $3/4$, respectively.

\section{Conclusion and Outlook}

In this work we introduced topological invariants and the concept of higher-order entanglement as new tools to characterize HOSPT phases.
First, we derived a topological many-body invariant, closely related to the discrete Wen-Zee response and fractional corner charges, which reveals the non-commutative algebra between flux operator and $C_n$ rotations. 
Second, we introduced the concept of `higher-order entanglement', to scrutinize and differentiate various higher-order topological phases from a hierarchical sequence of the entanglement structure.

It is expected that the concepts proposed in this paper can be generalized to a broad range of HOSPT phases in higher dimensions.  
For 3D HOSPT states, we expect that there exists a rotational Witten effect~\cite{chen2015anomalous,metlitski2013bosonic} where the magnetic monopole carries angular momentum so the topological invariant is defined via the non-commutative relation between the monopole insertion operator and the spatial rotation symmetry. 
%
%
Remarkably, such 3D HOSPT phases characterized by a rotational Witten effect, upon symmetry gauging, can potentially prompt a 3D $U(1)$ spin liquid with crystalline symmetry enriched monopole structure and intertwined coupling between the phonon mode and the emergent photon excitations! We anticipate that our results will enhance the search of new topological liquids in crystalline phases.
 
Another exciting direction is the application of our concepts to crystalline symmetry enriched quantum spin liquids and deconfined quantum critical points (DQCP)~\cite{song2018spinon,lee2019signatures,ning2019fractionalization,zou2018bulk}, where the crystalline symmetry interplays with the U(1) gauge field in a non-trivial way. 
In particular, in these systems the monopole operator also carries angular momentum and the instanton operator is odd under spatial rotation. Consequently, the instanton tunneling events are prohibited by spatial symmetry and the corresponding gauge theory is deconfined. We expect some exotic spatial symmetry enriched quantum spin liquids and DQCP can emerge after gauging the HOSPT (proximate HOSPT) system.
Such exploration also shed light on the search for crystalline symmetry enriched topological phases which can potentially host abundant and fascinating phenomenology. 

\textbf{Note}: When finishing this manuscript, we become aware of an unpublished work\cite{hughessc} related to this issue.

\section{Acknowledgments.}

The authors thank Izabella Lovas and Fabian Grusdt for stimulating discussions.
FP is funded by the European Research Council (ERC) under the European Unions Horizon 2020 research and innovation program (grant agreement No. 771537). FP acknowledges the support of the DFG Research Unit FOR 1807 through grants no.PO 1370/2-1, TRR80, and the Deutsche Forschungsgemeinschaft (DFG, German Research Foundation) under Germany's Excellence Strategy EXC-2111-390814868. This work is initiated at KITP and YY, FP are supported in part by the National Science Foundation under Grant No.NSF PHY- 1748958(KITP) during the Topological Quantum Matter program.


\begin{thebibliography}{73}%
\makeatletter
\providecommand \@ifxundefined [1]{%
 \@ifx{#1\undefined}
}%
\providecommand \@ifnum [1]{%
 \ifnum #1\expandafter \@firstoftwo
 \else \expandafter \@secondoftwo
 \fi
}%
\providecommand \@ifx [1]{%
 \ifx #1\expandafter \@firstoftwo
 \else \expandafter \@secondoftwo
 \fi
}%
\providecommand \natexlab [1]{#1}%
\providecommand \enquote  [1]{``#1''}%
\providecommand \bibnamefont  [1]{#1}%
\providecommand \bibfnamefont [1]{#1}%
\providecommand \citenamefont [1]{#1}%
\providecommand \href@noop [0]{\@secondoftwo}%
\providecommand \href [0]{\begingroup \@sanitize@url \@href}%
\providecommand \@href[1]{\@@startlink{#1}\@@href}%
\providecommand \@@href[1]{\endgroup#1\@@endlink}%
\providecommand \@sanitize@url [0]{\catcode `\\12\catcode `\$12\catcode
  `\&12\catcode `\#12\catcode `\^12\catcode `\_12\catcode `\%12\relax}%
\providecommand \@@startlink[1]{}%
\providecommand \@@endlink[0]{}%
\providecommand \url  [0]{\begingroup\@sanitize@url \@url }%
\providecommand \@url [1]{\endgroup\@href {#1}{\urlprefix }}%
\providecommand \urlprefix  [0]{URL }%
\providecommand \Eprint [0]{\href }%
\providecommand \doibase [0]{http://dx.doi.org/}%
\providecommand \selectlanguage [0]{\@gobble}%
\providecommand \bibinfo  [0]{\@secondoftwo}%
\providecommand \bibfield  [0]{\@secondoftwo}%
\providecommand \translation [1]{[#1]}%
\providecommand \BibitemOpen [0]{}%
\providecommand \bibitemStop [0]{}%
\providecommand \bibitemNoStop [0]{.\EOS\space}%
\providecommand \EOS [0]{\spacefactor3000\relax}%
\providecommand \BibitemShut  [1]{\csname bibitem#1\endcsname}%
\let\auto@bib@innerbib\@empty
\bibitem [{\citenamefont {{Pollmann}}\ \emph {et~al.}(2010)\citenamefont
  {{Pollmann}}, \citenamefont {{Turner}}, \citenamefont {{Berg}},\ and\
  \citenamefont {{Oshikawa}}}]{Pollmann2010}%
  \BibitemOpen
  \bibfield  {author} {\bibinfo {author} {\bibfnamefont {F.}~\bibnamefont
  {{Pollmann}}}, \bibinfo {author} {\bibfnamefont {A.~M.}\ \bibnamefont
  {{Turner}}}, \bibinfo {author} {\bibfnamefont {E.}~\bibnamefont {{Berg}}}, \
  and\ \bibinfo {author} {\bibfnamefont {M.}~\bibnamefont {{Oshikawa}}},\
  }\href@noop {} {\bibfield  {journal} {\bibinfo  {journal} {Phys. Rev. B}\
  }\textbf {\bibinfo {volume} {81}},\ \bibinfo {pages} {064439} (\bibinfo
  {year} {2010})}\BibitemShut {NoStop}%
\bibitem [{\citenamefont {Schuch}\ \emph {et~al.}(2011)\citenamefont {Schuch},
  \citenamefont {Perez-Garcia},\ and\ \citenamefont {Cirac}}]{Schuch2011-jx}%
  \BibitemOpen
  \bibfield  {author} {\bibinfo {author} {\bibfnamefont {N.}~\bibnamefont
  {Schuch}}, \bibinfo {author} {\bibfnamefont {D.}~\bibnamefont
  {Perez-Garcia}}, \ and\ \bibinfo {author} {\bibfnamefont {I.}~\bibnamefont
  {Cirac}},\ }\href@noop {} {\bibfield  {journal} {\bibinfo  {journal} {Phys.
  Rev. B}\ }\textbf {\bibinfo {volume} {84}},\ \bibinfo {pages} {165139}
  (\bibinfo {year} {2011})}\BibitemShut {NoStop}%
\bibitem [{\citenamefont {Chen}\ \emph {et~al.}(2011)\citenamefont {Chen},
  \citenamefont {Liu},\ and\ \citenamefont {Wen}}]{chen2011two}%
  \BibitemOpen
  \bibfield  {author} {\bibinfo {author} {\bibfnamefont {X.}~\bibnamefont
  {Chen}}, \bibinfo {author} {\bibfnamefont {Z.-X.}\ \bibnamefont {Liu}}, \
  and\ \bibinfo {author} {\bibfnamefont {X.-G.}\ \bibnamefont {Wen}},\
  }\href@noop {} {\bibfield  {journal} {\bibinfo  {journal} {Physical Review
  B}\ }\textbf {\bibinfo {volume} {84}},\ \bibinfo {pages} {235141} (\bibinfo
  {year} {2011})}\BibitemShut {NoStop}%
\bibitem [{\citenamefont {Senthil}(2015)}]{Senthil2015-tp}%
  \BibitemOpen
  \bibfield  {author} {\bibinfo {author} {\bibfnamefont {T.}~\bibnamefont
  {Senthil}},\ }\href@noop {} {\bibfield  {journal} {\bibinfo  {journal}
  {Annual Review of Condensed Matter Physics}\ }\textbf {\bibinfo {volume}
  {6}},\ \bibinfo {pages} {299} (\bibinfo {year} {2015})}\BibitemShut {NoStop}%
\bibitem [{\citenamefont {Yao}\ and\ \citenamefont
  {Kivelson}(2010)}]{yao2010fragile}%
  \BibitemOpen
  \bibfield  {author} {\bibinfo {author} {\bibfnamefont {H.}~\bibnamefont
  {Yao}}\ and\ \bibinfo {author} {\bibfnamefont {S.~A.}\ \bibnamefont
  {Kivelson}},\ }\href@noop {} {\bibfield  {journal} {\bibinfo  {journal}
  {Physical review letters}\ }\textbf {\bibinfo {volume} {105}},\ \bibinfo
  {pages} {166402} (\bibinfo {year} {2010})}\BibitemShut {NoStop}%
\bibitem [{\citenamefont {Fu}(2011)}]{fu2011topological}%
  \BibitemOpen
  \bibfield  {author} {\bibinfo {author} {\bibfnamefont {L.}~\bibnamefont
  {Fu}},\ }\href@noop {} {\bibfield  {journal} {\bibinfo  {journal} {Physical
  Review Letters}\ }\textbf {\bibinfo {volume} {106}},\ \bibinfo {pages}
  {106802} (\bibinfo {year} {2011})}\BibitemShut {NoStop}%
\bibitem [{\citenamefont {Hughes}\ \emph {et~al.}(2011)\citenamefont {Hughes},
  \citenamefont {Prodan},\ and\ \citenamefont {Bernevig}}]{hughes2011}%
  \BibitemOpen
  \bibfield  {author} {\bibinfo {author} {\bibfnamefont {T.~L.}\ \bibnamefont
  {Hughes}}, \bibinfo {author} {\bibfnamefont {E.}~\bibnamefont {Prodan}}, \
  and\ \bibinfo {author} {\bibfnamefont {B.~A.}\ \bibnamefont {Bernevig}},\
  }\href@noop {} {\bibfield  {journal} {\bibinfo  {journal} {Phys. Rev. B}\
  }\textbf {\bibinfo {volume} {83}},\ \bibinfo {pages} {245132} (\bibinfo
  {year} {2011})}\BibitemShut {NoStop}%
\bibitem [{\citenamefont {Hsieh}\ \emph {et~al.}(2012)\citenamefont {Hsieh},
  \citenamefont {Lin}, \citenamefont {Liu}, \citenamefont {Duan}, \citenamefont
  {Bansil},\ and\ \citenamefont {Fu}}]{hsieh2012topological}%
  \BibitemOpen
  \bibfield  {author} {\bibinfo {author} {\bibfnamefont {T.~H.}\ \bibnamefont
  {Hsieh}}, \bibinfo {author} {\bibfnamefont {H.}~\bibnamefont {Lin}}, \bibinfo
  {author} {\bibfnamefont {J.}~\bibnamefont {Liu}}, \bibinfo {author}
  {\bibfnamefont {W.}~\bibnamefont {Duan}}, \bibinfo {author} {\bibfnamefont
  {A.}~\bibnamefont {Bansil}}, \ and\ \bibinfo {author} {\bibfnamefont
  {L.}~\bibnamefont {Fu}},\ }\href@noop {} {\bibfield  {journal} {\bibinfo
  {journal} {Nature communications}\ }\textbf {\bibinfo {volume} {3}},\
  \bibinfo {pages} {982} (\bibinfo {year} {2012})}\BibitemShut {NoStop}%
\bibitem [{\citenamefont {Cheng}\ \emph {et~al.}(2016)\citenamefont {Cheng},
  \citenamefont {Zaletel}, \citenamefont {Barkeshli}, \citenamefont
  {Vishwanath},\ and\ \citenamefont {Bonderson}}]{cheng2016translational}%
  \BibitemOpen
  \bibfield  {author} {\bibinfo {author} {\bibfnamefont {M.}~\bibnamefont
  {Cheng}}, \bibinfo {author} {\bibfnamefont {M.}~\bibnamefont {Zaletel}},
  \bibinfo {author} {\bibfnamefont {M.}~\bibnamefont {Barkeshli}}, \bibinfo
  {author} {\bibfnamefont {A.}~\bibnamefont {Vishwanath}}, \ and\ \bibinfo
  {author} {\bibfnamefont {P.}~\bibnamefont {Bonderson}},\ }\href@noop {}
  {\bibfield  {journal} {\bibinfo  {journal} {Physical Review X}\ }\textbf
  {\bibinfo {volume} {6}},\ \bibinfo {pages} {041068} (\bibinfo {year}
  {2016})}\BibitemShut {NoStop}%
\bibitem [{\citenamefont {Ando}\ and\ \citenamefont
  {Fu}(2015)}]{ando2015topological}%
  \BibitemOpen
  \bibfield  {author} {\bibinfo {author} {\bibfnamefont {Y.}~\bibnamefont
  {Ando}}\ and\ \bibinfo {author} {\bibfnamefont {L.}~\bibnamefont {Fu}},\
  }\href@noop {} {\bibfield  {journal} {\bibinfo  {journal} {Annu. Rev.
  Condens. Matter Phys.}\ }\textbf {\bibinfo {volume} {6}},\ \bibinfo {pages}
  {361} (\bibinfo {year} {2015})}\BibitemShut {NoStop}%
\bibitem [{\citenamefont {Slager}\ \emph {et~al.}(2013)\citenamefont {Slager},
  \citenamefont {Mesaros}, \citenamefont {Juri{\v{c}}i{\'c}},\ and\
  \citenamefont {Zaanen}}]{slager2013space}%
  \BibitemOpen
  \bibfield  {author} {\bibinfo {author} {\bibfnamefont {R.-J.}\ \bibnamefont
  {Slager}}, \bibinfo {author} {\bibfnamefont {A.}~\bibnamefont {Mesaros}},
  \bibinfo {author} {\bibfnamefont {V.}~\bibnamefont {Juri{\v{c}}i{\'c}}}, \
  and\ \bibinfo {author} {\bibfnamefont {J.}~\bibnamefont {Zaanen}},\
  }\href@noop {} {\bibfield  {journal} {\bibinfo  {journal} {Nature Physics}\
  }\textbf {\bibinfo {volume} {9}},\ \bibinfo {pages} {98} (\bibinfo {year}
  {2013})}\BibitemShut {NoStop}%
\bibitem [{\citenamefont {Hong}\ and\ \citenamefont
  {Fu}(2017)}]{hong2017topological}%
  \BibitemOpen
  \bibfield  {author} {\bibinfo {author} {\bibfnamefont {S.}~\bibnamefont
  {Hong}}\ and\ \bibinfo {author} {\bibfnamefont {L.}~\bibnamefont {Fu}},\
  }\href@noop {} {\bibfield  {journal} {\bibinfo  {journal} {arXiv preprint
  arXiv:1707.02594}\ } (\bibinfo {year} {2017})}\BibitemShut {NoStop}%
\bibitem [{\citenamefont {Qi}\ and\ \citenamefont
  {Fu}(2015)}]{qi2015anomalous}%
  \BibitemOpen
  \bibfield  {author} {\bibinfo {author} {\bibfnamefont {Y.}~\bibnamefont
  {Qi}}\ and\ \bibinfo {author} {\bibfnamefont {L.}~\bibnamefont {Fu}},\
  }\href@noop {} {\bibfield  {journal} {\bibinfo  {journal} {Physical review
  letters}\ }\textbf {\bibinfo {volume} {115}},\ \bibinfo {pages} {236801}
  (\bibinfo {year} {2015})}\BibitemShut {NoStop}%
\bibitem [{\citenamefont {Huang}\ \emph {et~al.}(2017)\citenamefont {Huang},
  \citenamefont {Song}, \citenamefont {Huang},\ and\ \citenamefont
  {Hermele}}]{huang2017building}%
  \BibitemOpen
  \bibfield  {author} {\bibinfo {author} {\bibfnamefont {S.-J.}\ \bibnamefont
  {Huang}}, \bibinfo {author} {\bibfnamefont {H.}~\bibnamefont {Song}},
  \bibinfo {author} {\bibfnamefont {Y.-P.}\ \bibnamefont {Huang}}, \ and\
  \bibinfo {author} {\bibfnamefont {M.}~\bibnamefont {Hermele}},\ }\href@noop
  {} {\bibfield  {journal} {\bibinfo  {journal} {Physical Review B}\ }\textbf
  {\bibinfo {volume} {96}},\ \bibinfo {pages} {205106} (\bibinfo {year}
  {2017})}\BibitemShut {NoStop}%
\bibitem [{\citenamefont {Teo}\ and\ \citenamefont
  {Hughes}(2013)}]{teo2013existence}%
  \BibitemOpen
  \bibfield  {author} {\bibinfo {author} {\bibfnamefont {J.~C.}\ \bibnamefont
  {Teo}}\ and\ \bibinfo {author} {\bibfnamefont {T.~L.}\ \bibnamefont
  {Hughes}},\ }\href@noop {} {\bibfield  {journal} {\bibinfo  {journal}
  {Physical review letters}\ }\textbf {\bibinfo {volume} {111}},\ \bibinfo
  {pages} {047006} (\bibinfo {year} {2013})}\BibitemShut {NoStop}%
\bibitem [{\citenamefont {Song}\ \emph
  {et~al.}(2017{\natexlab{a}})\citenamefont {Song}, \citenamefont {Huang},
  \citenamefont {Fu},\ and\ \citenamefont {Hermele}}]{song2017topological}%
  \BibitemOpen
  \bibfield  {author} {\bibinfo {author} {\bibfnamefont {H.}~\bibnamefont
  {Song}}, \bibinfo {author} {\bibfnamefont {S.-J.}\ \bibnamefont {Huang}},
  \bibinfo {author} {\bibfnamefont {L.}~\bibnamefont {Fu}}, \ and\ \bibinfo
  {author} {\bibfnamefont {M.}~\bibnamefont {Hermele}},\ }\href@noop {}
  {\bibfield  {journal} {\bibinfo  {journal} {Physical Review X}\ }\textbf
  {\bibinfo {volume} {7}},\ \bibinfo {pages} {011020} (\bibinfo {year}
  {2017}{\natexlab{a}})}\BibitemShut {NoStop}%
\bibitem [{\citenamefont {Watanabe}\ \emph {et~al.}(2017)\citenamefont
  {Watanabe}, \citenamefont {Po},\ and\ \citenamefont
  {Vishwanath}}]{watanabe2017structure}%
  \BibitemOpen
  \bibfield  {author} {\bibinfo {author} {\bibfnamefont {H.}~\bibnamefont
  {Watanabe}}, \bibinfo {author} {\bibfnamefont {H.~C.}\ \bibnamefont {Po}}, \
  and\ \bibinfo {author} {\bibfnamefont {A.}~\bibnamefont {Vishwanath}},\
  }\href@noop {} {\bibfield  {journal} {\bibinfo  {journal} {arXiv preprint
  arXiv:1707.01903}\ } (\bibinfo {year} {2017})}\BibitemShut {NoStop}%
\bibitem [{\citenamefont {Po}\ \emph {et~al.}(2017)\citenamefont {Po},
  \citenamefont {Vishwanath},\ and\ \citenamefont {Watanabe}}]{po2017symmetry}%
  \BibitemOpen
  \bibfield  {author} {\bibinfo {author} {\bibfnamefont {H.~C.}\ \bibnamefont
  {Po}}, \bibinfo {author} {\bibfnamefont {A.}~\bibnamefont {Vishwanath}}, \
  and\ \bibinfo {author} {\bibfnamefont {H.}~\bibnamefont {Watanabe}},\
  }\href@noop {} {\bibfield  {journal} {\bibinfo  {journal} {Nature
  Communications}\ }\textbf {\bibinfo {volume} {8}},\ \bibinfo {pages} {50}
  (\bibinfo {year} {2017})}\BibitemShut {NoStop}%
\bibitem [{\citenamefont {Isobe}\ and\ \citenamefont
  {Fu}(2015)}]{isobe2015theory}%
  \BibitemOpen
  \bibfield  {author} {\bibinfo {author} {\bibfnamefont {H.}~\bibnamefont
  {Isobe}}\ and\ \bibinfo {author} {\bibfnamefont {L.}~\bibnamefont {Fu}},\
  }\href@noop {} {\bibfield  {journal} {\bibinfo  {journal} {Physical Review
  B}\ }\textbf {\bibinfo {volume} {92}},\ \bibinfo {pages} {081304} (\bibinfo
  {year} {2015})}\BibitemShut {NoStop}%
\bibitem [{\citenamefont {Benalcazar}\ \emph
  {et~al.}(2017{\natexlab{a}})\citenamefont {Benalcazar}, \citenamefont
  {Bernevig},\ and\ \citenamefont {Hughes}}]{benalcazar2017quantized}%
  \BibitemOpen
  \bibfield  {author} {\bibinfo {author} {\bibfnamefont {W.~A.}\ \bibnamefont
  {Benalcazar}}, \bibinfo {author} {\bibfnamefont {B.~A.}\ \bibnamefont
  {Bernevig}}, \ and\ \bibinfo {author} {\bibfnamefont {T.~L.}\ \bibnamefont
  {Hughes}},\ }\href@noop {} {\bibfield  {journal} {\bibinfo  {journal}
  {Science}\ }\textbf {\bibinfo {volume} {357}},\ \bibinfo {pages} {61}
  (\bibinfo {year} {2017}{\natexlab{a}})}\BibitemShut {NoStop}%
\bibitem [{\citenamefont {Benalcazar}\ \emph
  {et~al.}(2017{\natexlab{b}})\citenamefont {Benalcazar}, \citenamefont
  {Bernevig},\ and\ \citenamefont {Hughes}}]{benalcazar2017electric}%
  \BibitemOpen
  \bibfield  {author} {\bibinfo {author} {\bibfnamefont {W.~A.}\ \bibnamefont
  {Benalcazar}}, \bibinfo {author} {\bibfnamefont {B.~A.}\ \bibnamefont
  {Bernevig}}, \ and\ \bibinfo {author} {\bibfnamefont {T.~L.}\ \bibnamefont
  {Hughes}},\ }\href@noop {} {\bibfield  {journal} {\bibinfo  {journal}
  {Physical Review B}\ }\textbf {\bibinfo {volume} {96}},\ \bibinfo {pages}
  {245115} (\bibinfo {year} {2017}{\natexlab{b}})}\BibitemShut {NoStop}%
\bibitem [{\citenamefont {Schindler}\ \emph {et~al.}(2017)\citenamefont
  {Schindler}, \citenamefont {Cook}, \citenamefont {Vergniory}, \citenamefont
  {Wang}, \citenamefont {Parkin}, \citenamefont {Bernevig},\ and\ \citenamefont
  {Neupert}}]{schindler2017higher}%
  \BibitemOpen
  \bibfield  {author} {\bibinfo {author} {\bibfnamefont {F.}~\bibnamefont
  {Schindler}}, \bibinfo {author} {\bibfnamefont {A.~M.}\ \bibnamefont {Cook}},
  \bibinfo {author} {\bibfnamefont {M.~G.}\ \bibnamefont {Vergniory}}, \bibinfo
  {author} {\bibfnamefont {Z.}~\bibnamefont {Wang}}, \bibinfo {author}
  {\bibfnamefont {S.~S.}\ \bibnamefont {Parkin}}, \bibinfo {author}
  {\bibfnamefont {B.~A.}\ \bibnamefont {Bernevig}}, \ and\ \bibinfo {author}
  {\bibfnamefont {T.}~\bibnamefont {Neupert}},\ }\href@noop {} {\bibfield
  {journal} {\bibinfo  {journal} {arXiv preprint arXiv:1708.03636}\ } (\bibinfo
  {year} {2017})}\BibitemShut {NoStop}%
\bibitem [{\citenamefont {Langbehn}\ \emph {et~al.}(2017)\citenamefont
  {Langbehn}, \citenamefont {Peng}, \citenamefont {Trifunovic}, \citenamefont
  {von Oppen},\ and\ \citenamefont {Brouwer}}]{langbehn2017reflection}%
  \BibitemOpen
  \bibfield  {author} {\bibinfo {author} {\bibfnamefont {J.}~\bibnamefont
  {Langbehn}}, \bibinfo {author} {\bibfnamefont {Y.}~\bibnamefont {Peng}},
  \bibinfo {author} {\bibfnamefont {L.}~\bibnamefont {Trifunovic}}, \bibinfo
  {author} {\bibfnamefont {F.}~\bibnamefont {von Oppen}}, \ and\ \bibinfo
  {author} {\bibfnamefont {P.~W.}\ \bibnamefont {Brouwer}},\ }\href@noop {}
  {\bibfield  {journal} {\bibinfo  {journal} {Physical review letters}\
  }\textbf {\bibinfo {volume} {119}},\ \bibinfo {pages} {246401} (\bibinfo
  {year} {2017})}\BibitemShut {NoStop}%
\bibitem [{\citenamefont {Song}\ \emph
  {et~al.}(2017{\natexlab{b}})\citenamefont {Song}, \citenamefont {Fang},\ and\
  \citenamefont {Fang}}]{song2017d}%
  \BibitemOpen
  \bibfield  {author} {\bibinfo {author} {\bibfnamefont {Z.}~\bibnamefont
  {Song}}, \bibinfo {author} {\bibfnamefont {Z.}~\bibnamefont {Fang}}, \ and\
  \bibinfo {author} {\bibfnamefont {C.}~\bibnamefont {Fang}},\ }\href@noop {}
  {\bibfield  {journal} {\bibinfo  {journal} {Physical review letters}\
  }\textbf {\bibinfo {volume} {119}},\ \bibinfo {pages} {246402} (\bibinfo
  {year} {2017}{\natexlab{b}})}\BibitemShut {NoStop}%
\bibitem [{\citenamefont {Else}\ \emph {et~al.}(2019)\citenamefont {Else},
  \citenamefont {Po},\ and\ \citenamefont {Watanabe}}]{else2019fragile}%
  \BibitemOpen
  \bibfield  {author} {\bibinfo {author} {\bibfnamefont {D.~V.}\ \bibnamefont
  {Else}}, \bibinfo {author} {\bibfnamefont {H.~C.}\ \bibnamefont {Po}}, \ and\
  \bibinfo {author} {\bibfnamefont {H.}~\bibnamefont {Watanabe}},\ }\href@noop
  {} {\bibfield  {journal} {\bibinfo  {journal} {Physical Review B}\ }\textbf
  {\bibinfo {volume} {99}},\ \bibinfo {pages} {125122} (\bibinfo {year}
  {2019})}\BibitemShut {NoStop}%
\bibitem [{\citenamefont {Song}\ and\ \citenamefont
  {Schnyder}(2017)}]{song2017interaction}%
  \BibitemOpen
  \bibfield  {author} {\bibinfo {author} {\bibfnamefont {X.-Y.}\ \bibnamefont
  {Song}}\ and\ \bibinfo {author} {\bibfnamefont {A.~P.}\ \bibnamefont
  {Schnyder}},\ }\href@noop {} {\bibfield  {journal} {\bibinfo  {journal}
  {Physical Review B}\ }\textbf {\bibinfo {volume} {95}},\ \bibinfo {pages}
  {195108} (\bibinfo {year} {2017})}\BibitemShut {NoStop}%
\bibitem [{\citenamefont {You}\ \emph {et~al.}(2018{\natexlab{a}})\citenamefont
  {You}, \citenamefont {Litinski},\ and\ \citenamefont {von
  Oppen}}]{you2018higher}%
  \BibitemOpen
  \bibfield  {author} {\bibinfo {author} {\bibfnamefont {Y.}~\bibnamefont
  {You}}, \bibinfo {author} {\bibfnamefont {D.}~\bibnamefont {Litinski}}, \
  and\ \bibinfo {author} {\bibfnamefont {F.}~\bibnamefont {von Oppen}},\
  }\href@noop {} {\bibfield  {journal} {\bibinfo  {journal} {arXiv preprint
  arXiv:1810.10556}\ } (\bibinfo {year} {2018}{\natexlab{a}})}\BibitemShut
  {NoStop}%
\bibitem [{\citenamefont {Rasmussen}\ and\ \citenamefont
  {Lu}(2018{\natexlab{a}})}]{rasmussen2018intrinsically}%
  \BibitemOpen
  \bibfield  {author} {\bibinfo {author} {\bibfnamefont {A.}~\bibnamefont
  {Rasmussen}}\ and\ \bibinfo {author} {\bibfnamefont {Y.-M.}\ \bibnamefont
  {Lu}},\ }\href@noop {} {\bibfield  {journal} {\bibinfo  {journal} {arXiv
  preprint arXiv:1810.12317}\ } (\bibinfo {year}
  {2018}{\natexlab{a}})}\BibitemShut {NoStop}%
\bibitem [{\citenamefont {Rasmussen}\ and\ \citenamefont
  {Lu}(2018{\natexlab{b}})}]{rasmussen2018classification}%
  \BibitemOpen
  \bibfield  {author} {\bibinfo {author} {\bibfnamefont {A.}~\bibnamefont
  {Rasmussen}}\ and\ \bibinfo {author} {\bibfnamefont {Y.-M.}\ \bibnamefont
  {Lu}},\ }\href@noop {} {\bibfield  {journal} {\bibinfo  {journal} {arXiv
  preprint arXiv:1809.07325}\ } (\bibinfo {year}
  {2018}{\natexlab{b}})}\BibitemShut {NoStop}%
\bibitem [{\citenamefont {Thorngren}\ and\ \citenamefont
  {Else}(2018)}]{thorngren2018gauging}%
  \BibitemOpen
  \bibfield  {author} {\bibinfo {author} {\bibfnamefont {R.}~\bibnamefont
  {Thorngren}}\ and\ \bibinfo {author} {\bibfnamefont {D.~V.}\ \bibnamefont
  {Else}},\ }\href@noop {} {\bibfield  {journal} {\bibinfo  {journal} {Physical
  Review X}\ }\textbf {\bibinfo {volume} {8}},\ \bibinfo {pages} {011040}
  (\bibinfo {year} {2018})}\BibitemShut {NoStop}%
\bibitem [{\citenamefont {Benalcazar}\ \emph {et~al.}(2018)\citenamefont
  {Benalcazar}, \citenamefont {Li},\ and\ \citenamefont
  {Hughes}}]{benalcazar2018quantization}%
  \BibitemOpen
  \bibfield  {author} {\bibinfo {author} {\bibfnamefont {W.~A.}\ \bibnamefont
  {Benalcazar}}, \bibinfo {author} {\bibfnamefont {T.}~\bibnamefont {Li}}, \
  and\ \bibinfo {author} {\bibfnamefont {T.~L.}\ \bibnamefont {Hughes}},\
  }\href@noop {} {\bibfield  {journal} {\bibinfo  {journal} {arXiv preprint
  arXiv:1809.02142}\ } (\bibinfo {year} {2018})}\BibitemShut {NoStop}%
\bibitem [{\citenamefont {Zhang}\ \emph {et~al.}(2019)\citenamefont {Zhang},
  \citenamefont {Wang}, \citenamefont {Yang}, \citenamefont {Qi},\ and\
  \citenamefont {Gu}}]{zhang2019construction}%
  \BibitemOpen
  \bibfield  {author} {\bibinfo {author} {\bibfnamefont {J.-H.}\ \bibnamefont
  {Zhang}}, \bibinfo {author} {\bibfnamefont {Q.-R.}\ \bibnamefont {Wang}},
  \bibinfo {author} {\bibfnamefont {S.}~\bibnamefont {Yang}}, \bibinfo {author}
  {\bibfnamefont {Y.}~\bibnamefont {Qi}}, \ and\ \bibinfo {author}
  {\bibfnamefont {Z.-C.}\ \bibnamefont {Gu}},\ }\href@noop {} {\bibfield
  {journal} {\bibinfo  {journal} {arXiv preprint arXiv:1909.05519}\ } (\bibinfo
  {year} {2019})}\BibitemShut {NoStop}%
\bibitem [{\citenamefont {Tiwari}\ \emph {et~al.}(2019)\citenamefont {Tiwari},
  \citenamefont {Li}, \citenamefont {Bernevig}, \citenamefont {Neupert},\ and\
  \citenamefont {Parameswaran}}]{tiwari2019unhinging}%
  \BibitemOpen
  \bibfield  {author} {\bibinfo {author} {\bibfnamefont {A.}~\bibnamefont
  {Tiwari}}, \bibinfo {author} {\bibfnamefont {M.-H.}\ \bibnamefont {Li}},
  \bibinfo {author} {\bibfnamefont {B.}~\bibnamefont {Bernevig}}, \bibinfo
  {author} {\bibfnamefont {T.}~\bibnamefont {Neupert}}, \ and\ \bibinfo
  {author} {\bibfnamefont {S.}~\bibnamefont {Parameswaran}},\ }\href@noop {}
  {\bibfield  {journal} {\bibinfo  {journal} {arXiv preprint arXiv:1905.11421}\
  } (\bibinfo {year} {2019})}\BibitemShut {NoStop}%
\bibitem [{\citenamefont {Jiang}\ \emph {et~al.}(2019)\citenamefont {Jiang},
  \citenamefont {Cheng}, \citenamefont {Qi},\ and\ \citenamefont
  {Lu}}]{jiang2019generalized}%
  \BibitemOpen
  \bibfield  {author} {\bibinfo {author} {\bibfnamefont {S.}~\bibnamefont
  {Jiang}}, \bibinfo {author} {\bibfnamefont {M.}~\bibnamefont {Cheng}},
  \bibinfo {author} {\bibfnamefont {Y.}~\bibnamefont {Qi}}, \ and\ \bibinfo
  {author} {\bibfnamefont {Y.-M.}\ \bibnamefont {Lu}},\ }\href@noop {}
  {\bibfield  {journal} {\bibinfo  {journal} {arXiv preprint arXiv:1907.08596}\
  } (\bibinfo {year} {2019})}\BibitemShut {NoStop}%
\bibitem [{\citenamefont {You}\ \emph {et~al.}(2019)\citenamefont {You},
  \citenamefont {Burnell},\ and\ \citenamefont {Hughes}}]{you2019multipolar}%
  \BibitemOpen
  \bibfield  {author} {\bibinfo {author} {\bibfnamefont {Y.}~\bibnamefont
  {You}}, \bibinfo {author} {\bibfnamefont {F.}~\bibnamefont {Burnell}}, \ and\
  \bibinfo {author} {\bibfnamefont {T.~L.}\ \bibnamefont {Hughes}},\
  }\href@noop {} {\bibfield  {journal} {\bibinfo  {journal} {arXiv preprint
  arXiv:1909.05868}\ } (\bibinfo {year} {2019})}\BibitemShut {NoStop}%
\bibitem [{\citenamefont {Pollmann}\ \emph {et~al.}(2012)\citenamefont
  {Pollmann}, \citenamefont {Berg}, \citenamefont {Turner},\ and\ \citenamefont
  {Oshikawa}}]{pollmann2012symmetry}%
  \BibitemOpen
  \bibfield  {author} {\bibinfo {author} {\bibfnamefont {F.}~\bibnamefont
  {Pollmann}}, \bibinfo {author} {\bibfnamefont {E.}~\bibnamefont {Berg}},
  \bibinfo {author} {\bibfnamefont {A.~M.}\ \bibnamefont {Turner}}, \ and\
  \bibinfo {author} {\bibfnamefont {M.}~\bibnamefont {Oshikawa}},\ }\href@noop
  {} {\bibfield  {journal} {\bibinfo  {journal} {Physical review b}\ }\textbf
  {\bibinfo {volume} {85}},\ \bibinfo {pages} {075125} (\bibinfo {year}
  {2012})}\BibitemShut {NoStop}%
\bibitem [{\citenamefont {Liu}\ \emph {et~al.}(2019)\citenamefont {Liu},
  \citenamefont {Vishwanath},\ and\ \citenamefont {Khalaf}}]{liu2019shift}%
  \BibitemOpen
  \bibfield  {author} {\bibinfo {author} {\bibfnamefont {S.}~\bibnamefont
  {Liu}}, \bibinfo {author} {\bibfnamefont {A.}~\bibnamefont {Vishwanath}}, \
  and\ \bibinfo {author} {\bibfnamefont {E.}~\bibnamefont {Khalaf}},\
  }\href@noop {} {\bibfield  {journal} {\bibinfo  {journal} {Physical Review
  X}\ }\textbf {\bibinfo {volume} {9}},\ \bibinfo {pages} {031003} (\bibinfo
  {year} {2019})}\BibitemShut {NoStop}%
\bibitem [{\citenamefont {Araki}\ \emph {et~al.}(2019)\citenamefont {Araki},
  \citenamefont {Mizoguchi},\ and\ \citenamefont {Hatsugai}}]{araki2019mathbb}%
  \BibitemOpen
  \bibfield  {author} {\bibinfo {author} {\bibfnamefont {H.}~\bibnamefont
  {Araki}}, \bibinfo {author} {\bibfnamefont {T.}~\bibnamefont {Mizoguchi}}, \
  and\ \bibinfo {author} {\bibfnamefont {Y.}~\bibnamefont {Hatsugai}},\
  }\href@noop {} {\bibfield  {journal} {\bibinfo  {journal} {arXiv preprint
  arXiv:1906.00218}\ } (\bibinfo {year} {2019})}\BibitemShut {NoStop}%
\bibitem [{\citenamefont {Zhu}\ \emph {et~al.}(2019)\citenamefont {Zhu},
  \citenamefont {Loehr},\ and\ \citenamefont {Hughes}}]{zhu2019identifying}%
  \BibitemOpen
  \bibfield  {author} {\bibinfo {author} {\bibfnamefont {P.}~\bibnamefont
  {Zhu}}, \bibinfo {author} {\bibfnamefont {K.}~\bibnamefont {Loehr}}, \ and\
  \bibinfo {author} {\bibfnamefont {T.~L.}\ \bibnamefont {Hughes}},\
  }\href@noop {} {\bibfield  {journal} {\bibinfo  {journal} {arXiv preprint
  arXiv:1910.10180}\ } (\bibinfo {year} {2019})}\BibitemShut {NoStop}%
\bibitem [{\citenamefont {Wen}(1990)}]{wen1990topological}%
  \BibitemOpen
  \bibfield  {author} {\bibinfo {author} {\bibfnamefont {X.-G.}\ \bibnamefont
  {Wen}},\ }\href@noop {} {\bibfield  {journal} {\bibinfo  {journal}
  {International Journal of Modern Physics B}\ }\textbf {\bibinfo {volume}
  {4}},\ \bibinfo {pages} {239} (\bibinfo {year} {1990})}\BibitemShut {NoStop}%
\bibitem [{\citenamefont {Wen}\ and\ \citenamefont
  {Zee}(1998)}]{wen1998topological}%
  \BibitemOpen
  \bibfield  {author} {\bibinfo {author} {\bibfnamefont {X.-G.}\ \bibnamefont
  {Wen}}\ and\ \bibinfo {author} {\bibfnamefont {A.}~\bibnamefont {Zee}},\
  }\href@noop {} {\bibfield  {journal} {\bibinfo  {journal} {Physical Review
  B}\ }\textbf {\bibinfo {volume} {58}},\ \bibinfo {pages} {15717} (\bibinfo
  {year} {1998})}\BibitemShut {NoStop}%
\bibitem [{\citenamefont {You}\ \emph {et~al.}(2018{\natexlab{b}})\citenamefont
  {You}, \citenamefont {Devakul}, \citenamefont {Burnell},\ and\ \citenamefont
  {Neupert}}]{you2018highertitus}%
  \BibitemOpen
  \bibfield  {author} {\bibinfo {author} {\bibfnamefont {Y.}~\bibnamefont
  {You}}, \bibinfo {author} {\bibfnamefont {T.}~\bibnamefont {Devakul}},
  \bibinfo {author} {\bibfnamefont {F.~J.}\ \bibnamefont {Burnell}}, \ and\
  \bibinfo {author} {\bibfnamefont {T.}~\bibnamefont {Neupert}},\ }\href@noop
  {} {\bibfield  {journal} {\bibinfo  {journal} {Physical Review B}\ }\textbf
  {\bibinfo {volume} {98}},\ \bibinfo {pages} {235102} (\bibinfo {year}
  {2018}{\natexlab{b}})}\BibitemShut {NoStop}%
\bibitem [{\citenamefont {Han}\ \emph {et~al.}(2019)\citenamefont {Han},
  \citenamefont {Wang},\ and\ \citenamefont {Ye}}]{han2019generalized}%
  \BibitemOpen
  \bibfield  {author} {\bibinfo {author} {\bibfnamefont {B.}~\bibnamefont
  {Han}}, \bibinfo {author} {\bibfnamefont {H.}~\bibnamefont {Wang}}, \ and\
  \bibinfo {author} {\bibfnamefont {P.}~\bibnamefont {Ye}},\ }\href@noop {}
  {\bibfield  {journal} {\bibinfo  {journal} {Physical Review B}\ }\textbf
  {\bibinfo {volume} {99}},\ \bibinfo {pages} {205120} (\bibinfo {year}
  {2019})}\BibitemShut {NoStop}%
\bibitem [{\citenamefont {Else}\ and\ \citenamefont
  {Thorngren}(2019)}]{else2019topological}%
  \BibitemOpen
  \bibfield  {author} {\bibinfo {author} {\bibfnamefont {D.~V.}\ \bibnamefont
  {Else}}\ and\ \bibinfo {author} {\bibfnamefont {R.}~\bibnamefont
  {Thorngren}},\ }\href@noop {} {\bibfield  {journal} {\bibinfo  {journal}
  {arXiv preprint arXiv:1907.08204}\ } (\bibinfo {year} {2019})}\BibitemShut
  {NoStop}%
\bibitem [{\citenamefont {Li}\ and\ \citenamefont {Haldane}(2008)}]{Li-2008}%
  \BibitemOpen
  \bibfield  {author} {\bibinfo {author} {\bibfnamefont {H.}~\bibnamefont
  {Li}}\ and\ \bibinfo {author} {\bibfnamefont {F.~D.~M.}\ \bibnamefont
  {Haldane}},\ }\href {\doibase 10.1103/PhysRevLett.101.010504} {\bibfield
  {journal} {\bibinfo  {journal} {Phys. Rev. Lett.}\ }\textbf {\bibinfo
  {volume} {101}},\ \bibinfo {pages} {010504} (\bibinfo {year}
  {2008})}\BibitemShut {NoStop}%
\bibitem [{\citenamefont {Levin}\ and\ \citenamefont {Wen}(2006)}]{Levin-2006}%
  \BibitemOpen
  \bibfield  {author} {\bibinfo {author} {\bibfnamefont {M.}~\bibnamefont
  {Levin}}\ and\ \bibinfo {author} {\bibfnamefont {X.-G.}\ \bibnamefont
  {Wen}},\ }\href {\doibase 10.1103/PhysRevLett.96.110405} {\bibfield
  {journal} {\bibinfo  {journal} {Phys. Rev. Lett.}\ }\textbf {\bibinfo
  {volume} {96}},\ \bibinfo {eid} {110405} (\bibinfo {year}
  {2006})}\BibitemShut {NoStop}%
\bibitem [{\citenamefont {Kitaev}\ and\ \citenamefont
  {Preskill}(2006)}]{KitaevPreskill}%
  \BibitemOpen
  \bibfield  {author} {\bibinfo {author} {\bibfnamefont {A.}~\bibnamefont
  {Kitaev}}\ and\ \bibinfo {author} {\bibfnamefont {J.}~\bibnamefont
  {Preskill}},\ }\href {\doibase 10.1103/PhysRevLett.96.110404} {\bibfield
  {journal} {\bibinfo  {journal} {Phys. Rev. Lett.}\ }\textbf {\bibinfo
  {volume} {96}},\ \bibinfo {eid} {110404} (\bibinfo {year}
  {2006})}\BibitemShut {NoStop}%
\bibitem [{\citenamefont {White}(1992)}]{White:1992}%
  \BibitemOpen
  \bibfield  {author} {\bibinfo {author} {\bibfnamefont {S.~R.}\ \bibnamefont
  {White}},\ }\href {\doibase 10.1103/PhysRevLett.69.2863} {\bibfield
  {journal} {\bibinfo  {journal} {Phys. Rev. Lett.}\ }\textbf {\bibinfo
  {volume} {69}},\ \bibinfo {pages} {2863} (\bibinfo {year}
  {1992})}\BibitemShut {NoStop}%
\bibitem [{\citenamefont {Hauschild}\ and\ \citenamefont
  {Pollmann}(2018)}]{Hauschild2018}%
  \BibitemOpen
  \bibfield  {author} {\bibinfo {author} {\bibfnamefont {J.}~\bibnamefont
  {Hauschild}}\ and\ \bibinfo {author} {\bibfnamefont {F.}~\bibnamefont
  {Pollmann}},\ }\href {https://scipost.org/10.21468/SciPostPhysLectNotes.5}
  {\bibfield  {journal} {\bibinfo  {journal} {SciPost Phys. Lect. Notes}\ }
  (\bibinfo {year} {2018})}\BibitemShut {NoStop}%
\bibitem [{\citenamefont {Bibo}\ \emph {et~al.}(2019)\citenamefont {Bibo},
  \citenamefont {Lovas}, \citenamefont {You}, \citenamefont {Grusdt},\ and\
  \citenamefont {Pollmann}}]{bibo2019fractional}%
  \BibitemOpen
  \bibfield  {author} {\bibinfo {author} {\bibfnamefont {J.}~\bibnamefont
  {Bibo}}, \bibinfo {author} {\bibfnamefont {I.}~\bibnamefont {Lovas}},
  \bibinfo {author} {\bibfnamefont {Y.}~\bibnamefont {You}}, \bibinfo {author}
  {\bibfnamefont {F.}~\bibnamefont {Grusdt}}, \ and\ \bibinfo {author}
  {\bibfnamefont {F.}~\bibnamefont {Pollmann}},\ }\href@noop {} {\bibfield
  {journal} {\bibinfo  {journal} {arXiv preprint arXiv:1911.04149}\ } (\bibinfo
  {year} {2019})}\BibitemShut {NoStop}%
\bibitem [{\citenamefont {Kang}\ \emph {et~al.}(2018)\citenamefont {Kang},
  \citenamefont {Shiozaki},\ and\ \citenamefont {Cho}}]{kang2018many}%
  \BibitemOpen
  \bibfield  {author} {\bibinfo {author} {\bibfnamefont {B.}~\bibnamefont
  {Kang}}, \bibinfo {author} {\bibfnamefont {K.}~\bibnamefont {Shiozaki}}, \
  and\ \bibinfo {author} {\bibfnamefont {G.~Y.}\ \bibnamefont {Cho}},\
  }\href@noop {} {\bibfield  {journal} {\bibinfo  {journal} {arXiv preprint
  arXiv:1812.06999}\ } (\bibinfo {year} {2018})}\BibitemShut {NoStop}%
\bibitem [{\citenamefont {Wheeler}\ \emph {et~al.}(2018)\citenamefont
  {Wheeler}, \citenamefont {Wagner},\ and\ \citenamefont
  {Hughes}}]{wheeler2018many}%
  \BibitemOpen
  \bibfield  {author} {\bibinfo {author} {\bibfnamefont {W.~A.}\ \bibnamefont
  {Wheeler}}, \bibinfo {author} {\bibfnamefont {L.~K.}\ \bibnamefont {Wagner}},
  \ and\ \bibinfo {author} {\bibfnamefont {T.~L.}\ \bibnamefont {Hughes}},\
  }\href@noop {} {\bibfield  {journal} {\bibinfo  {journal} {arXiv preprint
  arXiv:1812.06990}\ } (\bibinfo {year} {2018})}\BibitemShut {NoStop}%
\bibitem [{\citenamefont {Wen}(2003)}]{wen2003quantum}%
  \BibitemOpen
  \bibfield  {author} {\bibinfo {author} {\bibfnamefont {X.-G.}\ \bibnamefont
  {Wen}},\ }\href@noop {} {\bibfield  {journal} {\bibinfo  {journal} {Physical
  review letters}\ }\textbf {\bibinfo {volume} {90}},\ \bibinfo {pages}
  {016803} (\bibinfo {year} {2003})}\BibitemShut {NoStop}%
\bibitem [{\citenamefont {Witten}(1991)}]{witten1991quantization}%
  \BibitemOpen
  \bibfield  {author} {\bibinfo {author} {\bibfnamefont {E.}~\bibnamefont
  {Witten}},\ }\href@noop {} {\bibfield  {journal} {\bibinfo  {journal}
  {Communications in Mathematical Physics}\ }\textbf {\bibinfo {volume}
  {137}},\ \bibinfo {pages} {29} (\bibinfo {year} {1991})}\BibitemShut
  {NoStop}%
\bibitem [{\citenamefont {Thorngren}\ and\ \citenamefont
  {Else}(2016)}]{thorngren2016gauging}%
  \BibitemOpen
  \bibfield  {author} {\bibinfo {author} {\bibfnamefont {R.}~\bibnamefont
  {Thorngren}}\ and\ \bibinfo {author} {\bibfnamefont {D.~V.}\ \bibnamefont
  {Else}},\ }\href@noop {} {\bibfield  {journal} {\bibinfo  {journal} {arXiv
  preprint arXiv:1612.00846}\ } (\bibinfo {year} {2016})}\BibitemShut {NoStop}%
\bibitem [{Note1()}]{Note1}%
  \BibitemOpen
  \bibinfo {note} {We can lift the corner degeneracy by adding a magnetic field
  at the corners resulting in fully gapped edges and corners, respectively.
  Subsequently, due to the plaquette entangled state at the center, each
  quadrant contains a fractional charge $l/4$ .}\BibitemShut {Stop}%
\bibitem [{\citenamefont {Elben}\ \emph {et~al.}(2019)\citenamefont {Elben},
  \citenamefont {Yu}, \citenamefont {Zhu}, \citenamefont {Hafezi},
  \citenamefont {Pollmann}, \citenamefont {Zoller},\ and\ \citenamefont
  {Vermersch}}]{elben2019many}%
  \BibitemOpen
  \bibfield  {author} {\bibinfo {author} {\bibfnamefont {A.}~\bibnamefont
  {Elben}}, \bibinfo {author} {\bibfnamefont {J.}~\bibnamefont {Yu}}, \bibinfo
  {author} {\bibfnamefont {G.}~\bibnamefont {Zhu}}, \bibinfo {author}
  {\bibfnamefont {M.}~\bibnamefont {Hafezi}}, \bibinfo {author} {\bibfnamefont
  {F.}~\bibnamefont {Pollmann}}, \bibinfo {author} {\bibfnamefont
  {P.}~\bibnamefont {Zoller}}, \ and\ \bibinfo {author} {\bibfnamefont
  {B.}~\bibnamefont {Vermersch}},\ }\href@noop {} {\bibfield  {journal}
  {\bibinfo  {journal} {arXiv preprint arXiv:1906.05011}\ } (\bibinfo {year}
  {2019})}\BibitemShut {NoStop}%
\bibitem [{\citenamefont {van Enk}\ and\ \citenamefont
  {Beenakker}(2012)}]{Enk2012}%
  \BibitemOpen
  \bibfield  {author} {\bibinfo {author} {\bibfnamefont {S.~J.}\ \bibnamefont
  {van Enk}}\ and\ \bibinfo {author} {\bibfnamefont {C.~W.~J.}\ \bibnamefont
  {Beenakker}},\ }\href {\doibase 10.1103/PhysRevLett.108.110503} {\bibfield
  {journal} {\bibinfo  {journal} {Phys. Rev. Lett.}\ }\textbf {\bibinfo
  {volume} {108}},\ \bibinfo {pages} {110503} (\bibinfo {year}
  {2012})}\BibitemShut {NoStop}%
\bibitem [{\citenamefont {Elben}\ \emph {et~al.}(2018)\citenamefont {Elben},
  \citenamefont {Vermersch}, \citenamefont {Dalmonte}, \citenamefont {Cirac},\
  and\ \citenamefont {Zoller}}]{Elben2018}%
  \BibitemOpen
  \bibfield  {author} {\bibinfo {author} {\bibfnamefont {A.}~\bibnamefont
  {Elben}}, \bibinfo {author} {\bibfnamefont {B.}~\bibnamefont {Vermersch}},
  \bibinfo {author} {\bibfnamefont {M.}~\bibnamefont {Dalmonte}}, \bibinfo
  {author} {\bibfnamefont {J.~I.}\ \bibnamefont {Cirac}}, \ and\ \bibinfo
  {author} {\bibfnamefont {P.}~\bibnamefont {Zoller}},\ }\href {\doibase
  10.1103/PhysRevLett.120.050406} {\bibfield  {journal} {\bibinfo  {journal}
  {Phys. Rev. Lett.}\ }\textbf {\bibinfo {volume} {120}},\ \bibinfo {pages}
  {050406} (\bibinfo {year} {2018})}\BibitemShut {NoStop}%
\bibitem [{\citenamefont {Amico}\ \emph {et~al.}(2008)\citenamefont {Amico},
  \citenamefont {Fazio}, \citenamefont {Osterloh},\ and\ \citenamefont
  {Vedral}}]{RevModPhys.80.517}%
  \BibitemOpen
  \bibfield  {author} {\bibinfo {author} {\bibfnamefont {L.}~\bibnamefont
  {Amico}}, \bibinfo {author} {\bibfnamefont {R.}~\bibnamefont {Fazio}},
  \bibinfo {author} {\bibfnamefont {A.}~\bibnamefont {Osterloh}}, \ and\
  \bibinfo {author} {\bibfnamefont {V.}~\bibnamefont {Vedral}},\ }\href
  {\doibase 10.1103/RevModPhys.80.517} {\bibfield  {journal} {\bibinfo
  {journal} {Rev. Mod. Phys.}\ }\textbf {\bibinfo {volume} {80}},\ \bibinfo
  {pages} {517} (\bibinfo {year} {2008})}\BibitemShut {NoStop}%
\bibitem [{\citenamefont {Peschel}\ and\ \citenamefont
  {Eisler}(2009)}]{Peschel_2009}%
  \BibitemOpen
  \bibfield  {author} {\bibinfo {author} {\bibfnamefont {I.}~\bibnamefont
  {Peschel}}\ and\ \bibinfo {author} {\bibfnamefont {V.}~\bibnamefont
  {Eisler}},\ }\href {\doibase 10.1088/1751-8113/42/50/504003} {\bibfield
  {journal} {\bibinfo  {journal} {Journal of Physics A: Mathematical and
  Theoretical}\ }\textbf {\bibinfo {volume} {42}},\ \bibinfo {pages} {504003}
  (\bibinfo {year} {2009})}\BibitemShut {NoStop}%
\bibitem [{\citenamefont {Alexandradinata}\ \emph {et~al.}(2011)\citenamefont
  {Alexandradinata}, \citenamefont {Hughes},\ and\ \citenamefont
  {Bernevig}}]{PhysRevB.84.195103}%
  \BibitemOpen
  \bibfield  {author} {\bibinfo {author} {\bibfnamefont {A.}~\bibnamefont
  {Alexandradinata}}, \bibinfo {author} {\bibfnamefont {T.~L.}\ \bibnamefont
  {Hughes}}, \ and\ \bibinfo {author} {\bibfnamefont {B.~A.}\ \bibnamefont
  {Bernevig}},\ }\href {\doibase 10.1103/PhysRevB.84.195103} {\bibfield
  {journal} {\bibinfo  {journal} {Phys. Rev. B}\ }\textbf {\bibinfo {volume}
  {84}},\ \bibinfo {pages} {195103} (\bibinfo {year} {2011})}\BibitemShut
  {NoStop}%
\bibitem [{\citenamefont {Prodan}\ \emph {et~al.}(2010)\citenamefont {Prodan},
  \citenamefont {Hughes},\ and\ \citenamefont
  {Bernevig}}]{PhysRevLett.105.115501}%
  \BibitemOpen
  \bibfield  {author} {\bibinfo {author} {\bibfnamefont {E.}~\bibnamefont
  {Prodan}}, \bibinfo {author} {\bibfnamefont {T.~L.}\ \bibnamefont {Hughes}},
  \ and\ \bibinfo {author} {\bibfnamefont {B.~A.}\ \bibnamefont {Bernevig}},\
  }\href {\doibase 10.1103/PhysRevLett.105.115501} {\bibfield  {journal}
  {\bibinfo  {journal} {Phys. Rev. Lett.}\ }\textbf {\bibinfo {volume} {105}},\
  \bibinfo {pages} {115501} (\bibinfo {year} {2010})}\BibitemShut {NoStop}%
\bibitem [{\citenamefont {Fidkowski}(2010)}]{PhysRevLett.104.130502}%
  \BibitemOpen
  \bibfield  {author} {\bibinfo {author} {\bibfnamefont {L.}~\bibnamefont
  {Fidkowski}},\ }\href {\doibase 10.1103/PhysRevLett.104.130502} {\bibfield
  {journal} {\bibinfo  {journal} {Phys. Rev. Lett.}\ }\textbf {\bibinfo
  {volume} {104}},\ \bibinfo {pages} {130502} (\bibinfo {year}
  {2010})}\BibitemShut {NoStop}%
\bibitem [{\citenamefont {Chandran}\ \emph {et~al.}(2014)\citenamefont
  {Chandran}, \citenamefont {Khemani},\ and\ \citenamefont
  {Sondhi}}]{chandran2014universal}%
  \BibitemOpen
  \bibfield  {author} {\bibinfo {author} {\bibfnamefont {A.}~\bibnamefont
  {Chandran}}, \bibinfo {author} {\bibfnamefont {V.}~\bibnamefont {Khemani}}, \
  and\ \bibinfo {author} {\bibfnamefont {S.~L.}\ \bibnamefont {Sondhi}},\
  }\href@noop {} {\bibfield  {journal} {\bibinfo  {journal} {Physical review
  letters}\ }\textbf {\bibinfo {volume} {113}},\ \bibinfo {pages} {060501}
  (\bibinfo {year} {2014})}\BibitemShut {NoStop}%
\bibitem [{\citenamefont {Zaletel}\ \emph {et~al.}(2013)\citenamefont
  {Zaletel}, \citenamefont {Mong},\ and\ \citenamefont
  {Pollmann}}]{PhysRevLett.110.236801}%
  \BibitemOpen
  \bibfield  {author} {\bibinfo {author} {\bibfnamefont {M.~P.}\ \bibnamefont
  {Zaletel}}, \bibinfo {author} {\bibfnamefont {R.~S.~K.}\ \bibnamefont
  {Mong}}, \ and\ \bibinfo {author} {\bibfnamefont {F.}~\bibnamefont
  {Pollmann}},\ }\href {\doibase 10.1103/PhysRevLett.110.236801} {\bibfield
  {journal} {\bibinfo  {journal} {Phys. Rev. Lett.}\ }\textbf {\bibinfo
  {volume} {110}},\ \bibinfo {pages} {236801} (\bibinfo {year}
  {2013})}\BibitemShut {NoStop}%
\bibitem [{\citenamefont {Chen}\ \emph {et~al.}(2015)\citenamefont {Chen},
  \citenamefont {Burnell}, \citenamefont {Vishwanath},\ and\ \citenamefont
  {Fidkowski}}]{chen2015anomalous}%
  \BibitemOpen
  \bibfield  {author} {\bibinfo {author} {\bibfnamefont {X.}~\bibnamefont
  {Chen}}, \bibinfo {author} {\bibfnamefont {F.~J.}\ \bibnamefont {Burnell}},
  \bibinfo {author} {\bibfnamefont {A.}~\bibnamefont {Vishwanath}}, \ and\
  \bibinfo {author} {\bibfnamefont {L.}~\bibnamefont {Fidkowski}},\ }\href@noop
  {} {\bibfield  {journal} {\bibinfo  {journal} {Physical Review X}\ }\textbf
  {\bibinfo {volume} {5}},\ \bibinfo {pages} {041013} (\bibinfo {year}
  {2015})}\BibitemShut {NoStop}%
\bibitem [{\citenamefont {Metlitski}\ \emph {et~al.}(2013)\citenamefont
  {Metlitski}, \citenamefont {Kane},\ and\ \citenamefont
  {Fisher}}]{metlitski2013bosonic}%
  \BibitemOpen
  \bibfield  {author} {\bibinfo {author} {\bibfnamefont {M.~A.}\ \bibnamefont
  {Metlitski}}, \bibinfo {author} {\bibfnamefont {C.}~\bibnamefont {Kane}}, \
  and\ \bibinfo {author} {\bibfnamefont {M.~P.}\ \bibnamefont {Fisher}},\
  }\href@noop {} {\bibfield  {journal} {\bibinfo  {journal} {Physical Review
  B}\ }\textbf {\bibinfo {volume} {88}},\ \bibinfo {pages} {035131} (\bibinfo
  {year} {2013})}\BibitemShut {NoStop}%
\bibitem [{\citenamefont {Song}\ \emph {et~al.}(2018)\citenamefont {Song},
  \citenamefont {He}, \citenamefont {Vishwanath},\ and\ \citenamefont
  {Wang}}]{song2018spinon}%
  \BibitemOpen
  \bibfield  {author} {\bibinfo {author} {\bibfnamefont {X.-Y.}\ \bibnamefont
  {Song}}, \bibinfo {author} {\bibfnamefont {Y.-C.}\ \bibnamefont {He}},
  \bibinfo {author} {\bibfnamefont {A.}~\bibnamefont {Vishwanath}}, \ and\
  \bibinfo {author} {\bibfnamefont {C.}~\bibnamefont {Wang}},\ }\href@noop {}
  {\bibfield  {journal} {\bibinfo  {journal} {arXiv preprint arXiv:1811.11182}\
  } (\bibinfo {year} {2018})}\BibitemShut {NoStop}%
\bibitem [{\citenamefont {Lee}\ \emph {et~al.}(2019)\citenamefont {Lee},
  \citenamefont {You}, \citenamefont {Sachdev},\ and\ \citenamefont
  {Vishwanath}}]{lee2019signatures}%
  \BibitemOpen
  \bibfield  {author} {\bibinfo {author} {\bibfnamefont {J.~Y.}\ \bibnamefont
  {Lee}}, \bibinfo {author} {\bibfnamefont {Y.-Z.}\ \bibnamefont {You}},
  \bibinfo {author} {\bibfnamefont {S.}~\bibnamefont {Sachdev}}, \ and\
  \bibinfo {author} {\bibfnamefont {A.}~\bibnamefont {Vishwanath}},\
  }\href@noop {} {\bibfield  {journal} {\bibinfo  {journal} {arXiv preprint
  arXiv:1904.07266}\ } (\bibinfo {year} {2019})}\BibitemShut {NoStop}%
\bibitem [{\citenamefont {Ning}\ \emph {et~al.}(2019)\citenamefont {Ning},
  \citenamefont {Zou},\ and\ \citenamefont
  {Cheng}}]{ning2019fractionalization}%
  \BibitemOpen
  \bibfield  {author} {\bibinfo {author} {\bibfnamefont {S.-Q.}\ \bibnamefont
  {Ning}}, \bibinfo {author} {\bibfnamefont {L.}~\bibnamefont {Zou}}, \ and\
  \bibinfo {author} {\bibfnamefont {M.}~\bibnamefont {Cheng}},\ }\href@noop {}
  {\bibfield  {journal} {\bibinfo  {journal} {arXiv preprint arXiv:1905.03276}\
  } (\bibinfo {year} {2019})}\BibitemShut {NoStop}%
\bibitem [{\citenamefont {Zou}(2018)}]{zou2018bulk}%
  \BibitemOpen
  \bibfield  {author} {\bibinfo {author} {\bibfnamefont {L.}~\bibnamefont
  {Zou}},\ }\href@noop {} {\bibfield  {journal} {\bibinfo  {journal} {Physical
  Review B}\ }\textbf {\bibinfo {volume} {97}},\ \bibinfo {pages} {045130}
  (\bibinfo {year} {2018})}\BibitemShut {NoStop}%
\bibitem [{\citenamefont {Dubinkin}\ and\ \citenamefont
  {Hughes}(2020)}]{hughessc}%
  \BibitemOpen
  \bibfield  {author} {\bibinfo {author} {\bibfnamefont {O.}~\bibnamefont
  {Dubinkin}}\ and\ \bibinfo {author} {\bibfnamefont {T.}~\bibnamefont
  {Hughes}},\ }\href@noop {} {\bibfield  {journal} {\bibinfo  {journal} {To
  Appear}\ } (\bibinfo {year} {2020})}\BibitemShut {NoStop}%
\end{thebibliography}
\end{document}